\documentclass[journal]{IEEEtran}
\usepackage{amsmath,amsfonts}
\usepackage{array}
\usepackage{textcomp}
\usepackage{stfloats}
\usepackage{url}
\usepackage{verbatim}
\usepackage{cite}
\usepackage{nameref}

\usepackage{enumitem}
\usepackage{stfloats}
\usepackage[pdftex]{graphicx}
\usepackage{amsthm}
\usepackage{amssymb}
\usepackage{authblk}
\usepackage{comment}
\usepackage{cite}
\usepackage[ruled, linesnumbered]{algorithm2e}
\usepackage{tabularx}
\usepackage{colortbl}
\usepackage[x11names]{xcolor}
\usepackage{multirow}
\usepackage[margin=0.75in]{geometry}
\usepackage{hyperref}
\usepackage[labelformat=parens,subrefformat=parens,caption=false]{subfig}
\usepackage[export]{adjustbox}
\usepackage{lipsum}  
\hypersetup{
    colorlinks=true,
    linkcolor=black,
    filecolor=magenta,      
    urlcolor=blue,
    citecolor =black,
}
\DeclareMathAlphabet{\mathdutchcal}{U}{dutchcal}{m}{n}
\SetMathAlphabet{\mathdutchcal}{bold}{U}{dutchcal}{b}{n}
\DeclareMathAlphabet{\mathdutchbcal}{U}{dutchcal}{b}{n}
\DeclareMathAlphabet{\mathdutchcal}{U}{eus}{b}{n}

\usepackage{booktabs}

\newtheorem{proposition}{Proposition}

\SetKw{KwInput}{Input}
\SetKw{KwOutput}{Output}
\SetKw{KwInitialize}{Initialize}
\SetNlSty{textbf}{}{:}
\SetKwProg{Fn}{Function}{:}{}
\SetKwRepeat{DoWhile}{do}{while}

\begin{document}

\title{Non-iterative Optimization of \\Trajectory and Radio Resource for Aerial Network}

\author{Hyeonsu Lyu, Jonggyu Jang, Harim Lee, and Hyun Jong Yang,}


\maketitle

\begin{abstract}
We address a joint trajectory planning, user association, resource allocation, and power control problem to maximize proportional fairness in the aerial IoT network, considering practical end-to-end quality-of-service (QoS) and communication schedules. Though the problem is rather ancient, apart from the fact that the previous approaches have never considered user- and time-specific QoS, we point out a prevalent mistake in coordinate optimization approaches adopted by the majority of the literature. Coordinate optimization approaches, which repetitively optimize radio resources for a fixed trajectory and vice versa, generally converge to local optima when all variables are differentiable. However, these methods often stagnate at a non-stationary point, significantly degrading the network utility in mixed-integer problems such as joint trajectory and radio resource optimization. We detour this problem by converting the formulated problem into the Markov decision process (MDP). Exploiting the beneficial characteristics of the MDP, we design a non-iterative framework that cooperatively optimizes trajectory and radio resources without initial trajectory choice. The proposed framework can incorporate various trajectory-planning algorithms such as the genetic algorithm, tree search, and reinforcement learning. Extensive comparisons with diverse baselines verify that the proposed framework significantly outperforms the state-of-the-art method, nearly achieving the global optimum.
Our implementation code is available at \href{https://github.com/hslyu/dbspf}{https://github.com/hslyu/dbspf}.
\end{abstract}

\section{Introduction}
\label{sec:Introduction}
Aerial networks have been considered a key enabler of the sixth-generation (6G) wireless communication.
ITU-R's suggestion of the key 6G usage scenarios highlights the role of aerial networks in achieving massive communications and ubiquitous connectivity \cite{IMT2030}.
This entails utilizing aerial vehicles for on-demand services and extensive coverage \cite{Kurt21-HAPS_survey}.

These use cases of aerial networks are closely aligned with current research focus on internet-of-things (IoT) networks, as documented in the 3GPP white papers \cite{3gpp.38.331, 3gpp.38.821}.
One of the most critical agendas for realizing aerial IoT networks involves communication scheduling to compensate for the constrained memory and energy capacity \cite{Dimitrios23_EnergyIoTCollection, Wenwen23-IoT}.

There is a lack of research concerning practical end-to-end requirements, including service scheduling and quality-of-service (QoS) requirements despite the proliferation of research on aerial networks. 
This is mainly because optimizing aerial networks becomes significantly intractable when such practical requirements are considered.
Communication scheduling entangles positional variables and physical-layer variables, such as user association (UA); frequency resource allocation (RA); and power control (PC) variables, along with multiple time steps.

Most research detours the variable coupling by adopting coordinate optimization approaches which reciprocally optimize  TP variables for fixed radio resource variables and vice versa \cite{Hao23-alternating_optimization}.
In the optimization-theoric viewpoint, these approaches are well-known to converge to local optima in differentiable problems, but may halt at the non-stationary point\footnote{A point is stationary if changes in any direction make utility degradation (e.g. local maxima). A point is non-stationary if the point is not stationary.} when optimizing mixed-integer problems, even if the objective is convex \cite{Tseng01-CoordinateDescent}.

\begin{figure}[tb]
    \centering
    \includegraphics[width=\columnwidth]{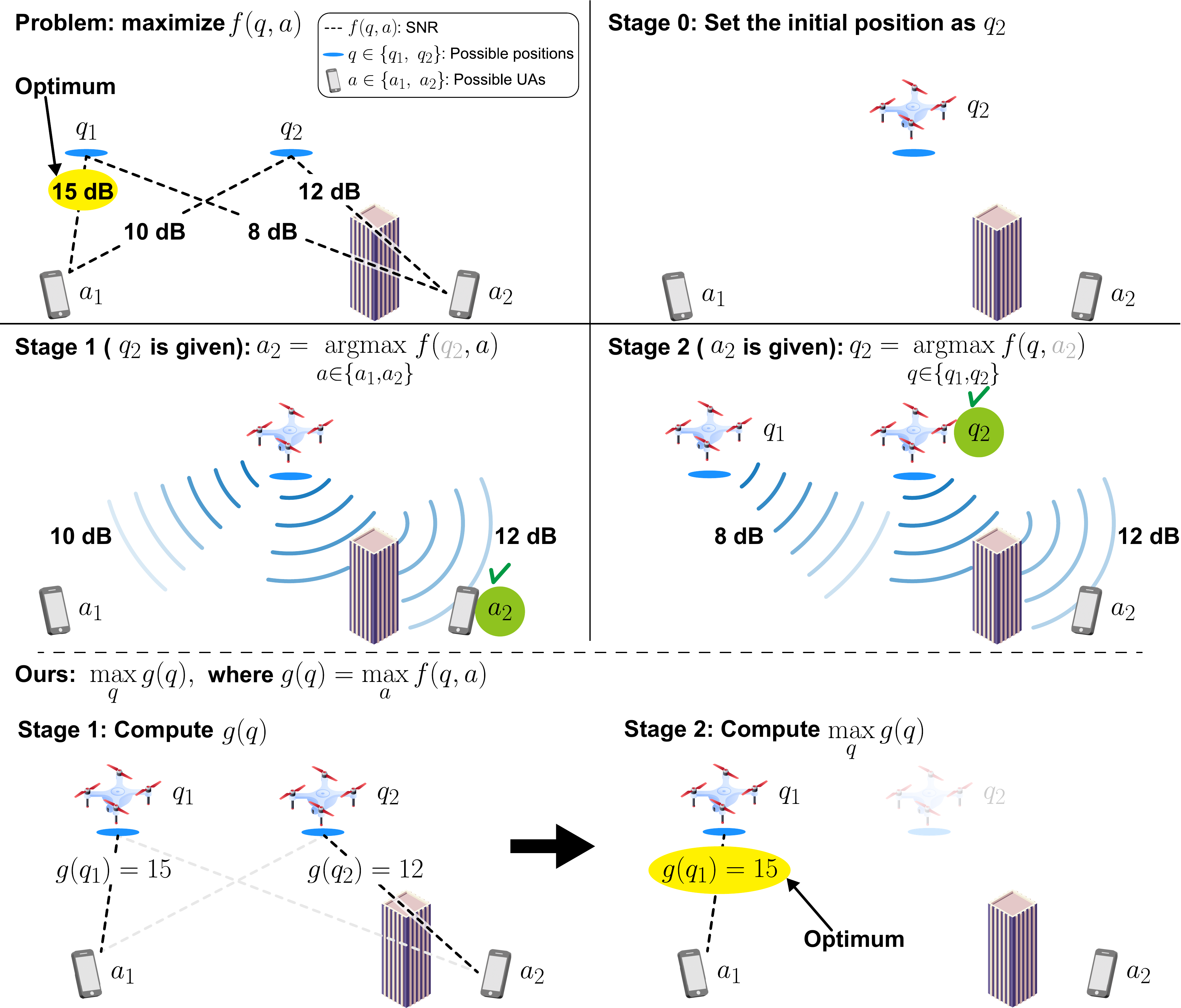}
    \caption{
    An illustrative failure scenario in coordinate optimization.
    Stages 0-2 demonstrate how coordinate optimization approaches converge to the non-stationary point.
    Once the initial $q$ is chosen (Stage 0), the following $a$ is accordingly determined (Stage 1).
    Then, the next iteration fails to find the optimal $q$ (Stage 2).
    Meanwhile, our approach adopts a hierarchical optimization that sweeps the entire variable space.
    }
    \label{fig:toy_example}
\end{figure}

We have found an interesting, but catastrophic interaction between trajectory-planning (TP) and radio resource management (RRM), where ``\textit{non-optimal initialization can trap the network variables in the non-stationary point}."
Figure~\ref{fig:toy_example} illustrates how the initialization deteriorates the network utility, unless the initial variable choice is optimal.
The discovery suggests that a surprising number of research works may inadvertently optimize the utility of aerial networks in a possibly incorrect manner
\cite{Zhang_FSO, Zeng_OFDMA_relay_GLOBECOM, Zeng_OFDMA_relay_TWC, Shen_multi_UAV_TP_PC_sumrate, Abbasi_TP_PC_sumrate_NOMA, Hu_AoI_MAC, Yang21-fig1_example, Gang21-fig1_example, Ying19-fig1_example, Nguyen22-fig1_example, Du23-fig1_example, Lu22-fig1_example, Liu23-fig1_example, Yang22-fig1_example, Yi22-fig1_example, Guangchi19-fig1_example}.

We ask ourselves \textit{``How can we optimize aerial networks while avoiding the curse of initialization?"} and find the answer in the non-iterative approach. 
As in Fig.~\ref{fig:toy_example}, we first convert radio resources as a function of trajectory.
Then, we find the optimal trajectory by integer programming.\footnote{We assume a discretized trajectory for the non-convex, non-differentiable problem. For a continuously differentiable trajectory, convex relaxations and optimizations can be applied.}
The optimal RRM is, in turn, determined by the trajectory, resulting in the optimal solution.

\begin{table}[bth]
    \centering
    \caption{Summary of the related works}
    \label{tab:related_works}
    \begin{tabularx}{\columnwidth}{cccccccc}
    \toprule
    Ref.$^*$ & 3D Traj.$^*$ & UA & RA & PC & QoS & Fairness & TCN$^*$  \\
    \cmidrule(r){1-1} \cmidrule(r{8pt}){2-8}
    \cite{Samir_TimeConstarined_IoT}  &   & \checkmark &  \checkmark &   &   &   &  \checkmark \\
    \arrayrulecolor{lightgray}
    \cmidrule(r){1-1} \cmidrule(r{10pt}){2-8}
    \cite{Al-Hilo_RIS_2d-TP, Zhu_TP_Recharge_station}   &   &  \checkmark &   &   &   &   &  \checkmark \\
    \cmidrule(r){1-1} \cmidrule(r{10pt}){2-8}
    \cite{Zhan_TP_WSN, Bayerlein_TP_sumrate_RL}   &  \checkmark &   &   &   &   &   &   \\
    \cmidrule(r){1-1} \cmidrule(r{10pt}){2-8}
    \cite{Shen_multi_UAV_TP_PC_sumrate}   &  \checkmark &   &   &  \checkmark &   &   &   \\
    \cmidrule(r){1-1} \cmidrule(r{10pt}){2-8}
    \cite{Abbasi_TP_PC_sumrate_NOMA}   &   &   &   &  \checkmark &   &   &   \\
    \cmidrule(r){1-1} \cmidrule(r{10pt}){2-8}
    \cite{Hu_AoI_MAC, Tong_DRL_TP_AoI}   &   &  \checkmark &   &   & \checkmark  &   &  \checkmark \\
    \cmidrule(r){1-1} \cmidrule(r{10pt}){2-8}
    \cite{You_TP_DataHarvesting}   &  \checkmark &  \checkmark &   &   &   &  \checkmark &   \\
    \cmidrule(r){1-1} \cmidrule(r{10pt}){2-8}
    \cite{Zhu_TP_DRL_Energy}   &   &  \checkmark &   &   &   &   &   \\
    \cmidrule(r){1-1} \cmidrule(r{10pt}){2-8}
    \cite{Liu_TP_PC_sumrate_RL}   & \checkmark  &   &   &  \checkmark &  \checkmark &   &   \\
    \cmidrule(r){1-1} \cmidrule(r{10pt}){2-8}
    \cite{Zeng_OFDMA_relay_GLOBECOM, Zeng_OFDMA_relay_TWC}   &  \checkmark &  \checkmark &  \checkmark &  \checkmark &  \checkmark &  \checkmark &   \\
    \cmidrule(r){1-1} \cmidrule(r{10pt}){2-8}
    Ours   &  \checkmark & \checkmark  &  \checkmark &  \checkmark & \checkmark  &  \checkmark &  \checkmark \\
    \arrayrulecolor{black}
    \bottomrule
    \end{tabularx}
    \\
    \rule{0pt}{9pt} * Ref.: Reference, Traj.: Trajectory, TCN: Time-critical networks.
\end{table}
While addressing the aforementioned research question, we additionally solve an open problem for the aerial IoT network. 
As highlighted in Table~\ref{tab:related_works},
there is a lack of research that studies proportional fairness (PF) maximization with practical end-to-end requirements.
Thus, we consider a joint TP; UA, RA, and PC problem for a single
UAV base station (UAV-BS) network where IoT devices punctually request a downlink stream above a certain QoS rate.
Appendix~\ref{Appendix:Related Works} provides an in-depth explanation of why the network scenario is considered, referring to adjacent research works.

We hierarchically optimize the PF maximization problem by subordinating the RRM to the TP.
Then, we can formulate the proposed problem as a Markov decision process (MDP) where various decision-making schemes can be applied.
For the RRM, we employ the Lagrangian method with Karush-Kuhn-Tucker (KKT) conditions, offering a low-complexity algorithm to accelerate the TP optimization.

\begin{figure}[htb]
    \centering
    \includegraphics[width=\columnwidth]{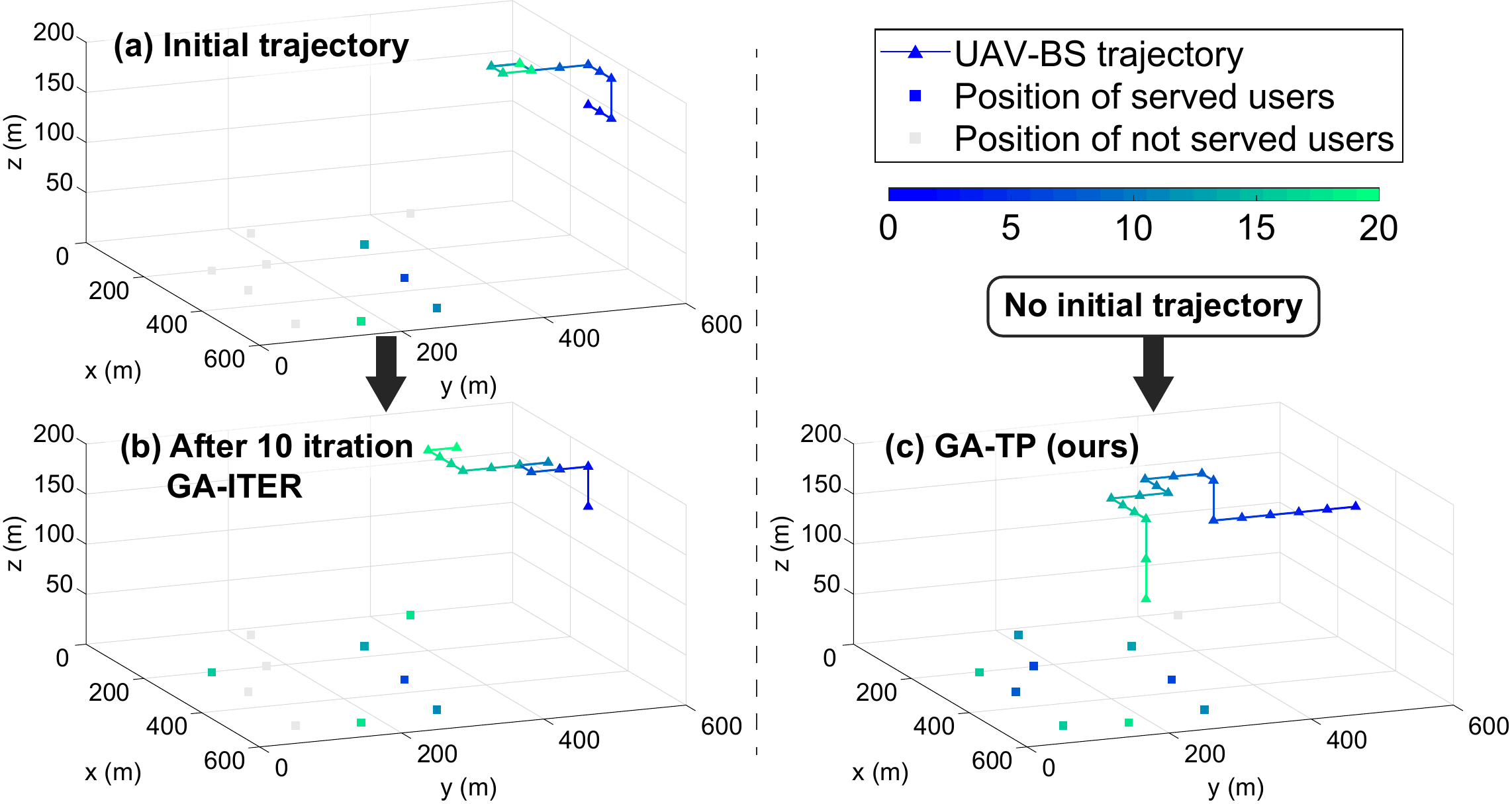}
    \caption{
    Motivated illustration of our contribution.
    Graphs (a) and (b) represent iterative optimization (GA-ITER), and graph (c) represents the proposed non-iterative approach (GA-TP).
    The two methods share the same environment configurations and algorithms except for the optimization order.
    While GA-ITER optimizes radio resources for a given initial trajectory, GA-TP adaptively optimizes radio resources according to the changing trajectory.
    Note that the initialization in (a) limits the changes of the updated trajectory, leading to a significant difference between the resulting trajectories in (b) and (c).}
    Detail descriptions are provided in the Sec.~\ref{sec:Simulation_Results}.
    \label{fig:contribution_visualization}
\end{figure}

The salient contributions of the paper can be summarized as follows.
\begin{itemize}
    \item 
    \textbf{Resolve the curse of initialization:}
    To the authors' knowledge, This research is the first to highlight that iterative optimization might be inherently unhelpful in aerial networks.
    We address the problem through the MDP reformulation, which provides a straightforward,  powerful non-iterative framework by separately optimizing TP and RRM problems.
    Figure~\ref{fig:contribution_visualization} briefly compares the result of the iterative optimization and the proposed method. 
    In Sec.~\ref{sec:Simulation_Results}, we show that the proposed method outperforms various comparison schemes and validate the robustness of the proposed method.
    \item 
    \textbf{Propose a novel temporal decoupling method:}
    We suggest a novel method to separate the PF problem into unit-time sub-problems.
    The PF entangles all controllable variables in the logarithm operations over the entire time steps. 
    Then, applying well-known RRM techniques, specialized to optimize unit-time network snapshots, becomes challenging.
    We devise multiple mathematical tricks to decompose the logarithm of sum-rate (PF formulation) into a summation of logarithms.
    Standing on the temporal decoupling, we show that the proposed RRM scheme tightly achieves the global optimum.
    \item 
    \textbf{Introduce a generalized water-filling algorithm:}
    We redesign the water-filling algorithm that can be applied to practical scenarios where QoS requirements exist.
    The conventional water-filling algorithm cannot find the UA and RA solution.
    The proposed algorithm automatically finds the optimal UA and RA combination when both the minimum resource requirements and resource budget co-exist.
    Numerical experiments show that the proposed method achieves the global maximum found by the genetic algorithm with the complexity of $\mathcal{O}(I^2)$ for $I$ users.
\end{itemize}

The remainder of the paper is organized as follows.
Section \ref{sec:System_model} introduces the system model of the IoT networks served by a UAV-BS and its mathematical representations.
Section \ref{sec:Problem formulation}-A formulates the PF maximization problem for TP, UA, RA, and PC, and Sec. \ref{sec:Problem formulation}-B decomposes the problem into the TP and RRM problems.
Section \ref{sec:Optimization of RRM and Control} presents in-depth explanations for the TP and RRM optimization.
Section \ref{sec:Simulation_Results} provides numerical evaluations of the proposed method.
Section \ref{sec:Conclusion} concludes the paper.

\section{System Model}
\label{sec:System_model}

\begin{figure}[htb]
    \centering
    \includegraphics[width=.95\columnwidth]{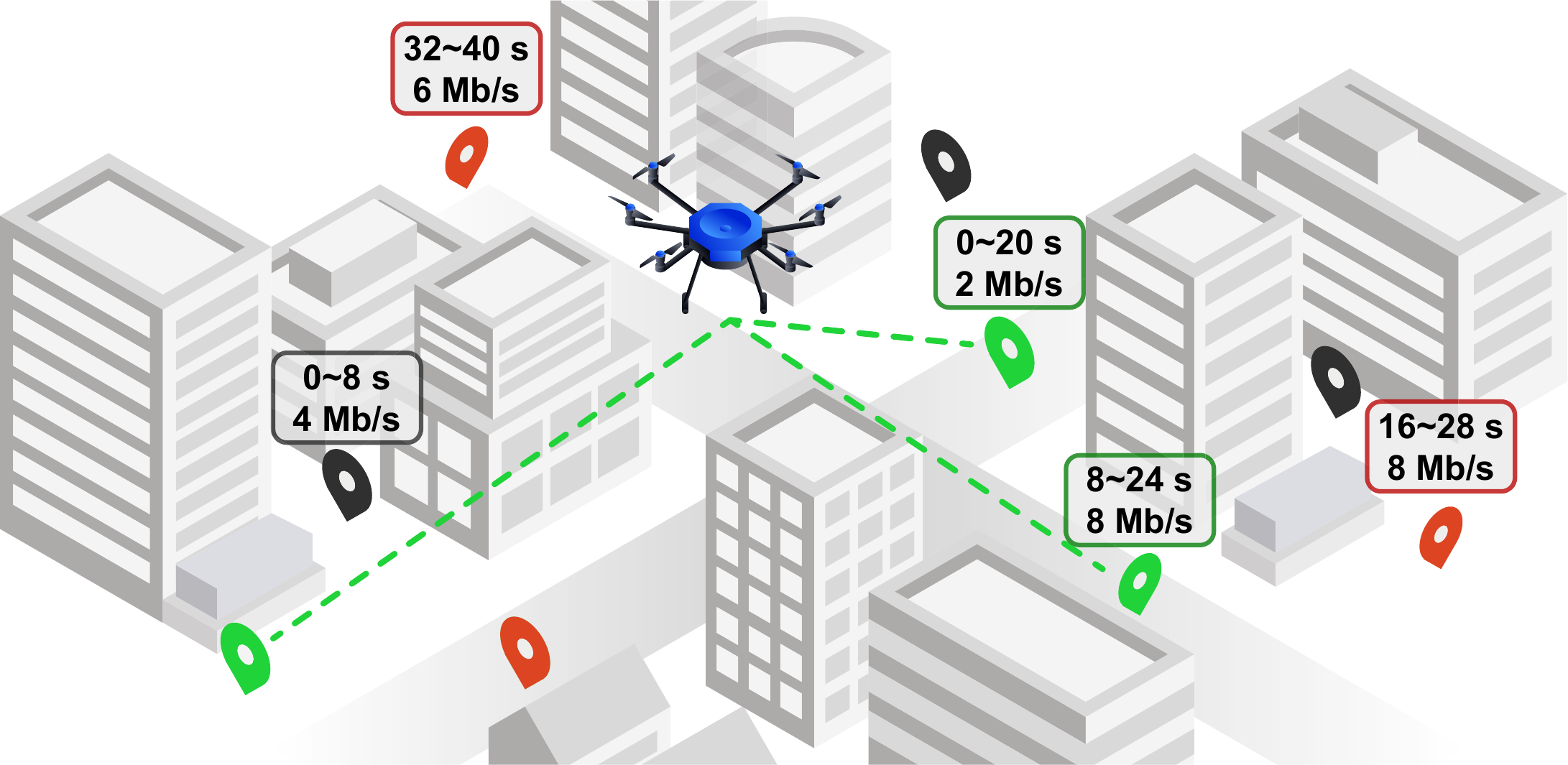}
    \caption{
    Illustration of the system model at time $t=20$.
    The dark gray marks represent users whose request periods have already expired.
    The green marks with dashed links represent users who are being serviced at the current time.
    The orange marks represent users whose request periods do not expire, but that are not currently being served.
    Users who have not expired may not be served if their request period does not start.
    The UAV-BS selects service users by jointly optimizing TP, UA, RA, and PC.
    }
    \label{fig:system_model}
\end{figure}

\subsection{Scenario Description}
We consider an IoT network where a single UAV-BS serves $I$ IoT devices for $T$ service time slots, as illustrated in Fig.~\ref{fig:system_model}.
In the network, the UAV-BS selectively provides downlink service to users due to the limitations in the available bandwidth $B$ and power $P$.
Each user has a downlink request period and QoS rate to satisfy its service requirements.
The users are indexed by the index set $\mathcal{I}=\{1,\dots,I\}$.

We assume that the flight map and service timeline are discretized.
The time slots are indexed by $\mathcal{T}=\{1,\dots,T\}$, each of which has the duration of $\Delta T$ (s).
The UAV-BS covers a square-shaped ground area with the size of $w\times w~\text{m}^2$, and
the flight map is represented by a set $\mathcal{Q}$ of three-dimensional grids discretized by an interval $\Delta Q$ (m) as follows:
\begin{align}
   \mathcal{Q} = \{ \Delta Q [i, j, k]^\mathrm{T} |& [i, j, k]^\mathrm{T} \in \mathbb{N}^3, h_{\text{min}} \leq k \Delta Q \leq h_{\text{max}}, \nonumber \\ 
   & 0 \leq i \Delta Q \leq w, 
   0 \leq j \Delta Q \leq w 
   \}, 
   \label{eq:map}
\end{align}
where $h_{\mathrm{min}}$ and $h_{\mathrm{max}}$ are the minimum and maximum altitude of the UAV-BS, respectively.

The UAV-BS is assumed to be located on a certain grid point of the flight map at the beginning of the $t$-th time slot.
The discrete trajectory---also known as waypoint-based navigation---has been a \textit{de facto standard} in a wide range of aerial vehicles, including commercial airplanes, copters, and fixed-wing vehicles \cite{Meier15-PX4, Ostroumov22-waypoint}.
We denote $\mathbf{q}^{(t)}$ as the position of the UAV-BS at the beginning of the $t$-th time slot, where $\mathbf{q}^{(t)}$ is defined as
\begin{align}
    \mathbf{q}^{(t)}=[x^{(t)}, y^{(t)}, h^{(t)}]^\mathrm{T} \in \mathcal{Q}.
    \label{eq:UAV-BS_position}
\end{align}
The initial position of the UAV-BS is denoted as $\mathbf{q}^{(0)}$.
The UAV-BS cannot exceed its maximum velocity $v$. 
In other words, the position vector $\mathbf{q}^{(t)}$ is constrained by 
\begin{align}
   \|\mathbf{q}^{(t)} - \mathbf{q}^{(t-1)}\|_2 \leq  v\Delta T,~\forall t \in \mathcal{T}.
   \label{eq:UAV-BS_velocity}
\end{align}
Combining \eqref{eq:UAV-BS_position} and \eqref{eq:UAV-BS_velocity}, all possible positions $\mathbf{q}^{(t)}$ at time slot $t$ are determined by the set $S(\mathbf{q}^{(t-1)}) = \{\mathbf{q}\in \mathcal{Q}|~ \|\mathbf{q} - \mathbf{q}^{(t-1)}\|_2 \leq v\Delta T \}$ as follows:
\begin{equation}
    \mathbf{q}^{(t)} \in S(\mathbf{q}^{(t-1)})
    \label{eq:tree_span}.
\end{equation}  

Users are stationary during the total service time.
The $i$-th user is located at $\mathbf{q}_i = [x_i, y_i, 0]^\mathrm{T}$.
To prolong the device lifetime, each user requests a downlink service only if the current flight time slot $t$ is within the short-term period\footnote{If $[s_i,T_i)=[0,T)$ for $i\in\mathcal{I}$, the proposed problem has a similar formulation with the problem in \cite{Zeng_OFDMA_relay_TWC} except for the objective function. However, the difference in the objective function causes a significant performance gap, which is shown in Sec.~\ref{sec:Simulation_Results}.} $[s_i, s_i+T_i)$. We define a binary indicator $d_i^{(t)}$ to specify user indices requesting services at time slot $t$ as follows: \begin{align}
    d_i^{(t)} = \begin{cases}
                1, &\text{if $s_i \leq t < s_i + T_i$} \\
                0, &\text{otherwise}.
                \end{cases}
\end{align}
For each time slot $t$, the UAV-BS determines a set of users to provide downlinks.
Accordingly, we define the UA variables as 
\begin{align}
    \alpha_i^{(t)} = \begin{cases}
                        1, & \hspace{-5pt} \text{if the UAV-BS serves user $i$ at time slot $t$} \\
                        0, & \hspace{-5pt} \text{otherwise}.
                     \end{cases}
\end{align}

We denote $\beta_i^{(t)}$ as the amount of frequency resources that the UAV-BS allocates to the $i$-th user at time slot $t$.
The summation of the allocated bandwidth proportions must not exceed the available bandwidth $B$, so we have the following constraint:
\begin{equation}
    \sum_{i \in \mathbf{A}^{(t)}} \beta_i^{(t)} \leq B,~\forall t \in \mathcal{T},
    \label{eq:sum_beta}
\end{equation}
where $\mathbf{A}^{(t)} = \{i \in \mathcal{I} | \alpha_i^{(t)}=1\}$ denotes the set of associated users at time slot $t$.

The UAV-BS regulates its transmission power by adjusting the power spectral density (PSD).
The variable $\rho_i^{(t)}$ is defined as the assigned PSD to the $i$-th user at time slot $t$. 
Denoting the maximum transmission power at each time slot as $P$, the transmission power is constrained as follows:
\begin{equation}
    \sum_{i \in \mathbf{A}^{(t)}} \rho_i^{(t)}\beta_i^{(t)} \leq P,~\forall t \in \mathcal{T}.
    \label{eq:sum_power}
\end{equation}

\subsection{Pathloss and Data Rate Model}

The propagation channel of the $i$-th user follows a probabilistic LoS channel \cite{Al-Hourani_A2G_channel}.
The probability that the $i$-th user has an LoS wireless link is defined as
\begin{align}
\label{eq:probability_LOS}
    \mathbb{P}_i^{(t)} = \Big(1+a\cdot\exp\big(-b(\theta_i^{(t)}-a)\big)\Big)^{-1},
\end{align}
where $\theta_i^{(t)} = \arcsin\left(\frac{h^{(t)}}{\|\mathbf{q}^{(t)}-\mathbf{q}_i\|_2}\right)$ is the elevation angle between the UAV-BS and the $i$-th user at time slot $t$. 
Variables $a$ and $b$ in \eqref{eq:probability_LOS} are environmental parameters that are determined according to the environmental characteristics \cite{Al-Hourani_A2G_channel, Holis_env_param}.
For example, possible pairs of ($a,b$) could be (4.76, 0.37) for a rural environment; and (9.64, 0.06) for a dense urban environment.
We remark that the $i$-th user has an NLoS link at time slot $t$ with a probability of $1-P_i^{(t)}$.

The average pathloss for the $i$-th user at time slot $t$ is defined as
\begin{align}
    \xi_i^{(t)}= \text{PL}(\mathbf{q}^{(t)}, \mathbf{q}_i)
    +\mathbb{P}_i^{(t)}\eta_{\text{LoS}}+(1-\mathbb{P}_i^{(t)})\eta_{\text{NLoS}},
    \label{eq:pathloss}
\end{align}
where pathloss $\text{PL}(\mathbf{q}^{(t)}, \mathbf{q}_i)=20\log\big(4\pi f$ $\|\mathbf{q}^{(t)}-\mathbf{q}_i\|_2/c\big)$; $f$ is the carrier frequency; $c$ is the speed of light; constant $\eta_{LoS}$ and $\eta_{NLoS}$ are the expected value of excessive pathloss for the LoS and NLoS links \cite{Al-Hourani_A2G_channel, A2GModel}.
The first term of \eqref{eq:pathloss} is the free-space pathloss and the sum of the remaining terms represents expected excessive pathloss \cite{Al-Hourani_A2G_channel}.

The data rate of the $i$-th user at time slot $t$ is defined as
\begin{align}
    R_i^{(t)} = d_i^{(t)}\beta_i^{(t)}\log_2\left(1+\frac{\rho_i^{(t)}10^{-\xi_i^{(t)}/10}}{N_0}\right),
    \label{eq:datarate}
\end{align}
where constant $N_0$ denotes additive white Gaussian noise PSD.

To ensure the served users' QoS requirements, the UAV-BS should serve user $i$ with a link that has a data rate higher than $r_i$ if the $i$-th user receives downlink data at time slot $t$.
Therefore, we have
\begin{align}
    R_i^{(t)} \geq \alpha_i^{(t)}r_i.
    \label{eq:QoS}
\end{align}
We define $R_i$ as the sum-rate of the $i$-th user across the UAV-BS service timeline, denoted as
\begin{align}
    R_i = \sum_{t\in \mathcal{T}} R_i^{(t)}.
\end{align}
Then, the PF of the served users is written by $\sum_{i\in\mathbf{A}}\log R_i$. 
We aim to maximize the PF that specializes in prioritized service scheduling. 
PF accounts for both the total network throughput and the number of served users as the logarithm operation prevents the network throughput from being concentrated on a few number of users.
However, sum-rate maximization prioritizes a user with the greatest spectral efficiency, not considering the fairness of service. 
Then the number of served users inevitably decreases when we target to maximize the sum-rate.

\section{Problem Formulation \label{sec:Problem formulation}}

\subsection{Original Full-Time Problem}
By gathering the previously defined constraints, the joint problem of TP, UA, RA, and PC is formulated to maximize the PF as 
\begin{subequations}
    \label{eq:p1}
    \begin{alignat}{3}
        & \bf{\mathdutchcal{P}1}: && \max_{\mathbf{Q,A,B,P}} &&  \sum_{i\in \mathbf{A}} \log R_i 
        \label{eq:p1_objective_function}\\
        &  &&       \textrm{~~~s.t.~~} && \mathbf{q}^{(t)} \in S(\mathbf{q}^{(t-1)}),
        \label{eq:p1_constraint_map} \\
        &  &&  &&   \beta_i^{(t)} \geq 0, \forall i\in\mathcal{I},
        \label{eq:p1_constraint_resource}\\
        &  &&  &&   \sum_{i \in \mathbf{A}^{(t)}} \beta_i^{(t)} \leq B,
        \label{eq:p1_constraint_sum_resource}\\
        &  &&  &&   \rho_i^{(t)} \geq 0, \forall i\in\mathcal{I},
        \label{eq:p1_constraint_psd}\\
        &  &&  &&   \sum_{i \in \mathbf{A}^{(t)}} \rho_i^{(t)}\beta_i^{(t)} \leq P,
        \label{eq:p1_constraint_power}\\
        &  &&  &&   R_i^{(t)}\geq \alpha_i^{(t)}r_i, \forall i\in\mathcal{I},
        \label{eq:p1_constraint_QoS} \\
        &  &&  &&   \forall t\in\mathcal{T} \text{ for \eqref{eq:p1_constraint_map}-\eqref{eq:p1_constraint_QoS}}. \nonumber
    \end{alignat}
\end{subequations}

For brevity of the notations, we define augmented vectors and matrices\footnote{
We define a vector-building operator $[e^{(t)}]^{t \in \mathcal{T}}$ to represent $[e^{(1)}, e^{(2)}, \dots, e^{(T)}]^{\rm{T}}$.
Then we define a matrix-building operator $[e_i^{(t)}]_{i\in I}^{t\in \mathcal{T}}$ as $[[e_1^{(t)}]^{t\in \mathcal{T}}, [e_2^{(t)}]^{t\in \mathcal{T}}, \dots, [e_I^{(t)}]^{t\in \mathcal{T}}]$.}
of variables as follows: 
$\mathbf{Q} = [\mathbf{q}^{(t)}]^{t\in \mathcal{T}}$,
$\mathbf{B} = [\beta_i^{(t)}]_{i\in \mathcal{I}}^{t\in \mathcal{T}}$ and
$\mathbf{P} = [\rho_i^{(t)}]_{i\in \mathcal{I}}^{t\in \mathcal{T}}$.
Also, $\mathbf{A} = \bigcup_{t\in\mathcal{T}} \mathbf{A}^{(t)}$ represents a set of users that have been served at least once during the UAV-BS service timeline.

Constraints \eqref{eq:p1_constraint_map} to \eqref{eq:p1_constraint_QoS} are described as follows:  
\eqref{eq:p1_constraint_map} correspond to \eqref{eq:tree_span};
\eqref{eq:p1_constraint_resource} and \eqref{eq:p1_constraint_psd} represent the non-negativity of the allocated bandwidth and PSD, respectively;
\eqref{eq:p1_constraint_sum_resource}, \eqref{eq:p1_constraint_power}, and \eqref{eq:p1_constraint_QoS} are equivalent with \eqref{eq:sum_beta}, \eqref{eq:sum_power}, and \eqref{eq:QoS}, respectively.

Problem $\bf{\mathdutchcal{P}1}$ is a non-convex mixed-integer problem that is generally known to be NP-hard \cite{Samuel_MINLP_NP_hard}, 
because the association variables $\alpha_i^{(t)}, \forall i,t$ are binary integers, and since the objective function is non-convex with respect to the position vector $\mathbf{q}^{(t)}$. 
Then, solving the problem within a feasible complexity becomes challenging if numerous time slots are involved in Problem $\bf{\mathdutchcal{P}1}$
because all the variables are coupled over the entire service timeline $\mathcal{T}$.
Therefore, we decompose Problem $\bf{\mathdutchcal{P}1}$ into $T$ sub-problems.
Each sub-problem has a Markov (or memoryless) property\cite{papoulis2002-probability}, making themselves independent of each other.

\subsection{On the Separation of Control and RRM}
\label{subsec:On the Separation of Control and RRM}
We first decompose Problem $\bf{\mathdutchcal{P}1}$ into sub-problems where each sub-problem corresponds to a single time slot. 
All controllable variables across time slots $t\in\mathcal{T}$ are closely coupled in the objective function of Problem $\bf{\mathdutchcal{P}1}$.
However, we can separate the variables and constraints related to the position from those related to the radio resources once Problem $\bf{\mathdutchcal{P}1}$ is temporally decomposed.

The key idea of the separation is converting summation into multiplication.
For example, $\Sigma_{i=1}^n i$ is equivalent with
\begin{align}
    \sum_{i=1}^n \hspace{-1pt} i \hspace{-1pt}=\hspace{-1pt} 1\cdot\frac{1+2}{1}\hspace{-1pt}\cdot\hspace{-1pt}\frac{1+2+3}{1+2}\hspace{-1pt}\cdots\hspace{-1pt}\frac{1+\cdots + n}{1+\cdots + n-1}.\hspace{-3pt}
\end{align}
Using the trick, we convert $\log R_i = \log \big(\sum_{t\in\mathcal{T}}R_i^{(t)}\big)$ into $\log R_i = \sum_{t\in\mathcal{T}}\log \big(1+R_i^{(t)}/\sum_{k=1}^{t-1}R_i^{(k)}\big)$.
Then, we can define a lower-bound of Problem $\bf{\mathdutchcal{P}1}$ which consists of $T$ sub-problems.

\begin{proposition}[Lower bound of Problem $\bf{\mathdutchcal{P}1}$]
    \label{prop1:1}
    \textit{The following inequality holds:}
    \begin{flalign}
        &\max_{\mathbf{Q},\mathbf{A}, \mathbf{B},\mathbf{P}} \sum_{i\in \mathbf{A}} \log R_i \\
        &=
        \max_{\mathbf{Q}} \max_{\mathbf{A}, \mathbf{B},\mathbf{P}} \sum_{t\in\mathcal{T}}\sum_{i\in \mathbf{A}} \log \left(1 + \frac{R_i^{(t)}}{\sum_{k= 0}^{t-1} R_i^{(k)}}\right) \\
        &\geq
        \max_{\mathbf{Q}} \hspace{-1pt} \sum_{t\in\mathcal{T}} \max_{\mathbf{A}^{(t)}, \mathbf{B}^{(t)}, \mathbf{P}^{(t)}} \hspace{-5pt}
        \sum_{i\in \mathbf{A}^{(t)}} \hspace{-1pt} \log \hspace{-2pt}\left(1 + \frac{R_i^{(t)}}{\sum_{k=0}^{t-1} R_i^{(k)}}\right)\hspace{-1pt},
        \label{eq:pt_lower_bound}
    \end{flalign}
    where $R_i^{(0)}=1$ for all $i\in\mathcal{I}$, 
    $\mathbf{A}^{(t)} = \{i \in \mathcal{I} | \alpha_i^{(t)}=1\}$,
    $\mathbf{B}^{(t)} = [\beta_i^{(t)}]_{i\in \mathcal{I}}$, and
    $\mathbf{P}^{(t)} = [\rho_i^{(t)}]_{i\in \mathcal{I}}$.
\end{proposition}
\begin{IEEEproof}
    The proof is shown in Appendix \ref{appendix:prop1}.
\end{IEEEproof}
From Proposition \ref{prop1:1}, we relax Problem $\bf{\mathdutchcal{P}1}$ to maximize the lower bound of the problem; that is, Eq.~\eqref{eq:pt_lower_bound} can be re-written as
\begin{equation}
    \mathbf{\mathdutchcal{P}2}: \max_{\substack{
            \mathbf{q}^{(t)} \in S(\mathbf{q}^{(t-1)}), \\
            t\in\mathcal{T}}
         }
    \sum_{t=1}^T f\big(\mathbf{q}^{(t)}\big), 
    \label{eq:MDP_objective}
\end{equation}
where
\begin{subequations}
    \begin{align}
        f(\mathbf{q}^{(t)}) = &
        \max_{\mathbf{A}^{(t)}, \mathbf{B}^{(t)},\mathbf{P}^{(t)}} 
        \sum_{i\in \mathbf{A}^{(t)}} \log \left(1 + \frac{R_i^{(t)}}{\sum_{k= 0}^{t-1} R_i^{(k)}}\right) \\
        & \text{~~~~~~s.t.~~~~}  \eqref{eq:p1_constraint_resource}-\eqref{eq:p1_constraint_QoS}.
    \end{align} 
    \label{eq:lookahead_value}
\end{subequations}
Here, Problem $\bf{\mathdutchcal{P}2}$ only involves with the positional variable $\mathbf{q}^{(t)}$ and Eq.~\eqref{eq:lookahead_value} associates with the radio resource variables $\mathbf{A}^{(t)}$, $\mathbf{B}^{(t)}$, and $\mathbf{P}^{(t)}$.
We remark that computing $f(\mathbf{q}^{(t)})$ requires $R_i^{(k)}$ for $k=1,...,t-1$.
Thus, $f(\mathbf{q}^{(t)})$ need to be sequentially computed $f(\mathbf{q}^{(t)})$ from $t=1$ to $t=T$.

\begin{figure}[htb]
    \centering
    \includegraphics[width=\columnwidth]{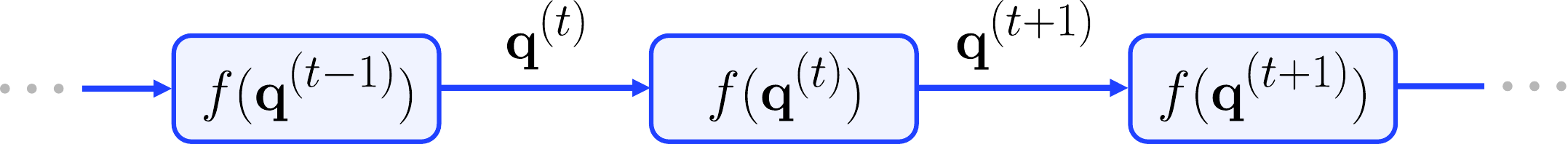}
    \caption{
    Illustration of the objective in problem $\bf{\mathdutchcal{P}2}$ in the viewpoint of MDP.
    For each time slot $t$, state is given as all variables before $t$; action is $\mathbf{q}^{(t)}$; and reward is $f(\mathbf{q}^{(t)})$.
    }
    \label{fig:Markov representation}
\end{figure}

The computing process can be considered a Markov decision process (MDP) in that sequential actions $\mathbf{q}^{(1)},...,\mathbf{q}^{(T)}$ determine a reward as $\sum_{t\in\mathcal{T}}f(\mathbf{q}^{(t)})$.
Once the optimal RRM scheme $f(\mathbf{q}^{(t)})$ is given, 
the control problem $\bf{\mathdutchcal{P}2}$  can be considered as finding the best solution of the MDP, as depicted in Fig.~\ref{fig:Markov representation}.

\section{Optimization of RRM and Control}
\label{sec:Optimization of RRM and Control}
The MDP reward $f(\mathbf{q}^{(t)})$ is ideally designed if we can find the optimal UA, RA, and PC.
Through the optimization-theoric approach, we develop a fast and accurate RRM that makes various decision-making schemes affordable in terms of the computing resource budget.
Then, the global optimum of Problem $\bf{\mathdutchcal{P}2}$ can be obtained by finding the optimal trajectory $\mathbf{q}^{(1)},...,\mathbf{q}^{(T)}$.
We aim to find the solution of Problem $\bf{\mathdutchcal{P}2}$ by applying various decision-making schemes to find the optimal trajectory.

\subsection{Finding the Optimal Radio Resource Management}

We first optimize the integer variable $\mathbf{A}^{(t)}$ by suggesting a generalized water-filling algorithm that can jointly optimize $\mathbf{A}^{(t)}$ and $\mathbf{B}^{(t)}$.
Then, the RA and PC variables, $\mathbf{B}^{(t)}$ and $\mathbf{P}^{(t)}$, are iteratively optimized by using the Lagrangian method and the KKT conditions.

\subsubsection{Initial User Association}
To determine optimal $\mathbf{A}^{(t)}$, 
we first assume that the PSD variables $\rho_i^{(t)}$ are equally assigned as $\rho_i^{(t)} = P/B$ for all $i\in \mathcal{I}$. 
Then the problem \eqref{eq:lookahead_value} in $f(\mathbf{q})$ can be simplified as 
\begin{subequations}
    \begin{alignat}{3}
        &\max_{\mathbf{A}^{(t)}, \mathbf{B}^{(t)}} && 
        \sum_{i\in \mathcal{I}} \log \left(1 + \frac{R_i^{(t)}}{\sum_{k= 0}^{t-1} R_i^{(k)}}\right) 
        \label{eq:p_UA_RA_objective} \\
        & \text{~~~s.t.}
          && \beta_i^{(t)} \geq \alpha_i^{(t)}r_i/e_i^{(t)}, \forall i, 
        \label{eq:p_UA_RA_beta}\\
        & && \sum_{i \in \mathbf{A}^{(t)}} \beta_i^{(t)} \leq B, 
        \label{eq:p_UA_RA_sum_beta}
    \end{alignat}
    \label{eq:p_UA_RA}
\end{subequations}
where $e_i^{(t)} = \log_2\big(1+\frac{\rho_i^{(t)}10^{-\xi_i^{(t)}/10}}{N_0}\big)$ is the spectral efficiency of the $i$-th user at time slot $t$.
This can be considered as a water-filling problem with non-zero resource requirements $\alpha_i^{(t)}r_i/e_i^{(t)}$.
Therefore, the optimal $\mathbf{B}^{(t)}$ is also provided as a water-filling solution.

\begin{figure}[hbt]
    \centering
    \includegraphics[width=.85\columnwidth]{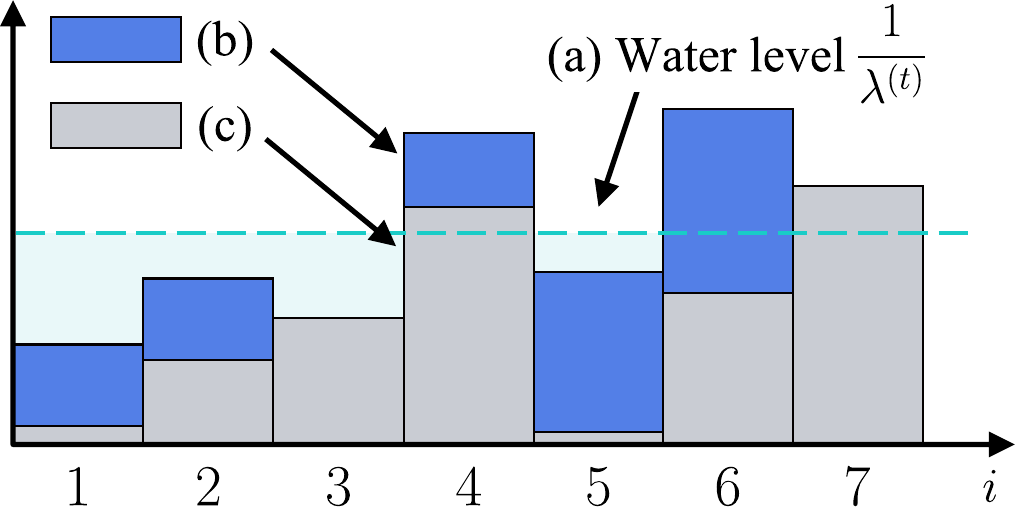}
    \caption{
    Visualization of the water-filling solution $\beta_i^{(t)}$ in \eqref{eq:beta_given_alpha} for 7 users.
    (a) The dashed line implies the water level. 
    (b) The blue column indicates the minimum RA requirement $r_i/e_i^{(t)}$. 
    (c) The grey column represents the given ground heights
    $1/w_i^{(t)}=\sum_{k=0}^{t-1}R_i^{(k)}/e_i^{(t)}$, which implies that the UAV-BS allocates a low bandwidth proportion to a user if the user has a high total received data and low spectral efficiency.
    }
    \label{fig:WF}
\end{figure}

We can obtain the optimal $\mathbf{B}^{(t)}$ by using the KKT conditions.
For a given $\mathbf{A}^{(t)}$ that does not violate constraint \eqref{eq:p_UA_RA_beta} and \eqref{eq:p_UA_RA_sum_beta},
the optimal $\beta_i^{(t)}$ is
\begin{align}
   \beta_i^{(t)} = 
   \begin{cases}
        \max\bigg(\frac{r_i}{e_i^{(t)}}, \frac{1}{\lambda^{(t)}}-\frac{1}{w_i^{(t)}}\bigg) & \text{, if $\alpha_i^{(t)}d_i^{(t)}=1$,} \\
        ~~~~~~~~~~~~~~~~~~ 0 & \text{, otherwise,}
   \end{cases}
   \label{eq:beta_given_alpha}
\end{align}
where $\lambda^{(t)}$ can be uniquely determined to satisfy the constraint \eqref{eq:p_UA_RA_sum_beta} with equality condition\footnote{We use the bisection method to find the solution for $\lambda$.} and $1/w_i^{(t)}=\sum_{k=0}^{t-1}R_i^{(k)}/e_i^{(t)}$.
Physically, the solution implies that the UAV-BS allocates a little frequency resource to the user who has a high total received data and low spectral efficiency.
The derivation of \eqref{eq:beta_given_alpha} is provided in Appendix \ref{appendix:derive_beta}.
Figure~\ref{fig:WF} depicts our intuition on the lower-bounded water-filling solution.

\begin{algorithm}[tb]
    \caption{User Association Algorithm}
    \label{alg:init_ua_ra}
    \KwInput{$\mathbf{q}^{(t)}$, $\mathbf{P}^{(t)}$}\\
    \KwOutput{$\mathbf{A}^{(t)}$, $\mathbf{B}^{(t)}$}\\
    $\mathbf{A}_{\rm{current}}^{(t)} \leftarrow \{\}$. \\
    \While{$\mathbf{A}_{\rm{current}}^{(t)} \neq \mathbf{A}_{\rm{next}}^{(t)}$}{
        $\mathbf{A}_{\rm{current}}^{(t)} \leftarrow \mathbf{A}_{\rm{next}}^{(t)}$ \\
        $\mathcal{I}_{\rm{feasible}} \leftarrow \{i\in\mathcal{I} | \sum_{i \in \mathbf{A}_{\mathrm{current}+i}^{(t)}} r_i/e_i^{(t)} \leq B \}$ \\
        $\mathbf{A}_{\rm{next}}^{(t)}\leftarrow$ The best UA set among $\mathbf{A}_{\rm{current}}^{(t)}$ and $\mathbf{A}_{\mathrm{current}+i}^{(t)} ~\text{for}~i\in\mathcal{I}_{\rm{feasible}}$.\\
    }
    $\mathbf{A}^{(t)}\leftarrow \mathbf{A}_{\rm{next}}^{(t)}$ \\
    Compute $\mathbf{B}^{(t)}$ for $\mathbf{A}^{(t)}$ by using \eqref{eq:beta_given_alpha}.
\end{algorithm}

Now, the objective \eqref{eq:p_UA_RA_objective} can be maximized by finding the optimal $\mathbf{A}^{(t)}$ as 
the optimal $\mathbf{B}^{(t)}$ can be obtained for arbitrary $\mathbf{A}^{(t)}$,
we adopt an incremental algorithm that finds a local-optimal solution (Alg.~\ref{alg:init_ua_ra}). 
The key idea of the algorithm is to sequentially find a better UA combination than the current UA combination $\mathbf{A}_{\mathrm{current}}^{(t)}$.

Let $\mathbf{A}_{\mathrm{current}}^{(t)}$ be a UA set and $\mathbf{A}_{\mathrm{current}+i}^{(t)}$ contains one more user not in $\mathbf{A}_{\mathrm{current}}^{(t)}$, denoted as
\begin{equation}
    \mathbf{A}_{\mathrm{current}+i}^{(t)} = \mathbf{A}_{\mathrm{current}}^{(t)}\cup \{i\}.
\end{equation}

If $\mathbf{A}_{\mathrm{current}+i}^{(t)}$ satisfies constraint \eqref{eq:p_UA_RA_beta} and \eqref{eq:p_UA_RA_sum_beta},
$\mathbf{A}_{\text{current}+i}^{(t)}$ could be a maximizer of Problem \eqref{eq:p_UA_RA}.
Two constraints \eqref{eq:p_UA_RA_beta} and \eqref{eq:p_UA_RA_sum_beta} can be simplified,
so we can define a set of feasible additive users as
\begin{equation}
    \mathcal{I}_{\mathrm{feasible}} = \big\{i\in\mathcal{I} | \Sigma_{i \in \mathbf{A}_{\mathrm{current}+i}^{(t)}} r_i/e_i^{(t)} \leq B \big\}.
    \label{eq:feasible_UA_set}
\end{equation}

Then, we can define a new set $\mathbf{A}_{\text{next}}^{(t)}$ that makes the objective \eqref{eq:p_UA_RA_objective} be the greatest among the set 
\vspace{0.05cm}
$\mathbf{A}_{\mathrm{current}}^{(t)}$ and $\mathbf{A}_{\mathrm{current}+i}^{(t)}$ for $i\in\mathcal{I}_{\mathrm{feasible}}$.
\vspace{.05cm}
If $\mathbf{A}_{\mathrm{next}}^{(t)}$ is equal to $\mathbf{A}_{\mathrm{current}}^{(t)}$, set $\mathbf{A}_{\mathrm{next}}^{(t)}$ is a solution.
Otherwise, $\mathbf{A}_{\mathrm{current}}^{(t)}$ is updated to $\mathbf{A}_{\mathrm{next}}^{(t)}$ and the above procedure is repeated until finding the solution.

By iteratively updating the UA set from the empty set $\mathbf{A}_{\mathrm{current}}^{(t)} = \{\}$, the proposed algorithm can find the local-optimal UA and RA variables.

\subsubsection{Resource Allocation and Power Control}
The optimal $\mathbf{B}^{(t)}$ and $\mathbf{P}^{(t)}$ for the given $\mathbf{A}^{(t)}$ is optimized by iteratively optimizing the following two problems.

Once the initial $\mathbf{A}^{(t)}$ and $\mathbf{B}^{(t)}$ is determined by Alg.~\ref{alg:init_ua_ra}, $\mathbf{P}^{(t)}$ can be accordingly determined by solving 
\begin{subequations}
    \begin{align}
        \max_{\mathbf{P}^{(t)}} &
        \sum_{i\in \mathbf{A}^{(t)}} \log \left(1 + \frac{R_i^{(t)}}{\sum_{k= 0}^{t-1} R_i^{(k)}}\right) \\
        & \text{~~s.t~~}  \eqref{eq:p1_constraint_power},
        \eqref{eq:p1_constraint_psd},
        \text{ and } \eqref{eq:p1_constraint_QoS}.
    \end{align} 
    \label{eq:rrm_ra}
\end{subequations}
Similarly, fixing $\mathbf{A}^{(t)}$ and $\mathbf{P}^{(t)}$ gives
\begin{subequations}
    \begin{align}
        \max_{\mathbf{B}^{(t)}} &
        \sum_{i\in \mathbf{A}^{(t)}} \log \left(1 + \frac{R_i^{(t)}}{\sum_{k= 0}^{t-1} R_i^{(k)}}\right) \\
        & \text{~~s.t.~~}  \eqref{eq:p1_constraint_resource},
        \eqref{eq:p1_constraint_sum_resource},
        \eqref{eq:p1_constraint_QoS},
        ~\text{and } \eqref{eq:p1_constraint_power}.
    \end{align} 
    \label{eq:rrm_psd}
\end{subequations}
Using the convexity of these two problems, 

The Lagrangian dual method and KKT conditions are applied to minimize the computing time,
and we find an $\mathcal{O}(I^2)$ algorithm (Alg.~3 in Appendix~\ref{Appendix:Optimal Resource Allocation}) and closed-form solution (Eq.~(69) in Appendix~\ref{Appendix:Optimal Power Control}) for Problem \eqref{eq:rrm_ra} and \eqref{eq:rrm_psd}, respectively.
Appendices~\ref{Appendix:Optimal Resource Allocation} and \ref{Appendix:Optimal Power Control} describe in-depth optimization processes.

\begin{algorithm}[tb]
    \caption{Radio Resource Management $f(\cdot)$}
    \label{alg:RRM}
    \KwInput{$\mathbf{q}^{(t)}$}\\
    \KwOutput{$f(\mathbf{q}^{(t)})$, $\mathbf{A}^{(t)}$, $\mathbf{B}^{(t)}$, $\mathbf{P}^{(t)}$}\\
    \KwInitialize{$f \leftarrow 0$, $f_{\mathrm{prev}} \leftarrow \infty$, $\rho_i^{(t)} = P/B, \forall i\in \mathcal{I}$} \\
    \Fn{$f(\mathbf{q}^{(t)})$}{
        Update $\mathbf{A}^{(t)}$ and $\mathbf{B}^{(t)}$ by using Alg. \ref{alg:init_ua_ra}. \\
        \While{$\|f-f_{\mathrm{prev}} \|>\epsilon$}{
            Update $\mathbf{B}^{(t)}$ by using Alg. 3 in Appendix~\ref{Appendix:Optimal Resource Allocation}. \\
            Update $\mathbf{P}^{(t)}$ by using (69) in Appendix~\ref{Appendix:Optimal Power Control}.\\
            $f_{\mathrm{prev}} \leftarrow f$,~
            $f\leftarrow \sum_{i\in \mathbf{A}^{(t)}} \log \bigg(1 + \frac{R_i^{(t)}}{\sum_{k= 0}^{t-1} R_i^{(k)}}\bigg)$\\
        }
        $\mathbf{return}~ f$ \\
    }
\end{algorithm}

\begin{figure*}[htb]
    \centering
    \includegraphics[width=\textwidth]{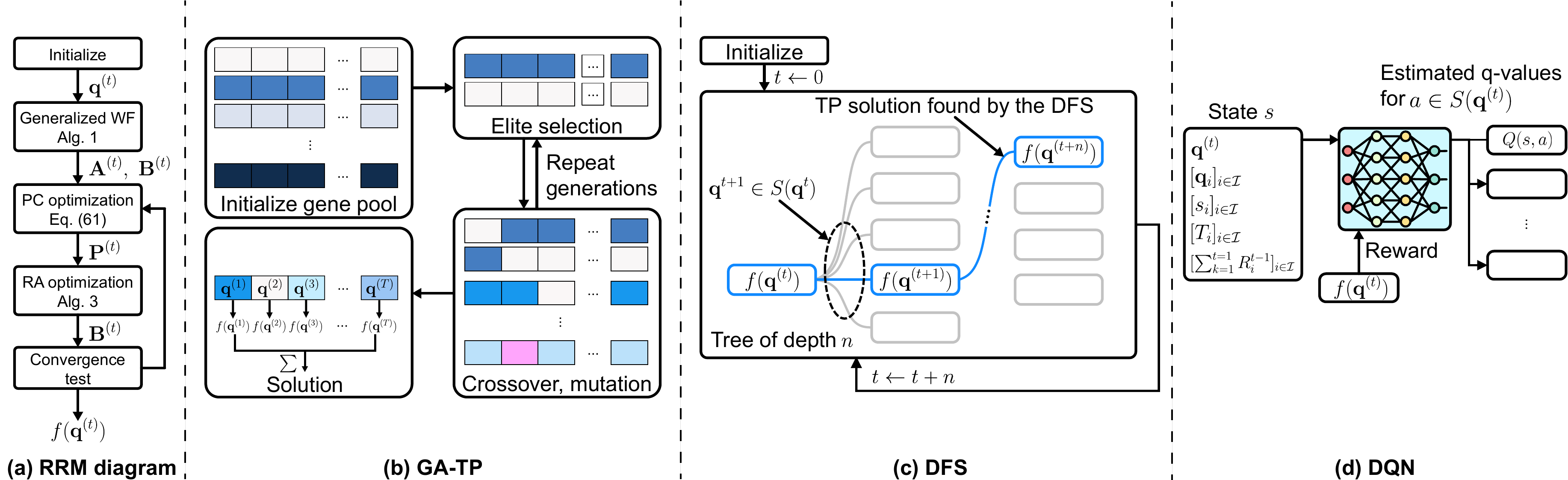}
    \caption{
    Visualization of the proposed algorithms
    (a) Algorithm flowchart of the RRM scheme in Alg.~\ref{alg:RRM}. (b) Visualization of GA-TP. Each gene corresponds to a trajectory. Fitness for each gene is computed at the elite selection. (c) Visualization of DFS with a tree representation of length $n$ sub-trajectory. (d) State, reward, and action for DQN.}
    \label{fig:TP-RRM-visualization}
\end{figure*}

\subsubsection{Overall RRM Algorithm}
Combining the UA, RA, and PC schemes altogether, the RRM scheme to compute $f(\mathbf{q^{(t)}})$ can be summarized as Alg.~\ref{alg:RRM}.
We first initialize the PC variables as $\rho_i^{(t)} = P/B$ for all $i\in\mathcal{I}$.
Then, in Line 5, we obtain initial UA and RA variables for given $\mathbf{P}^{(t)}$ by Alg.~\ref{alg:init_ua_ra}.
After the UA variables are initialized, the RA and PC variables are iteratively optimized in Lines 6 to 9, until the objective function converges.
Figure~\ref{fig:TP-RRM-visualization}a visualizes these processes as algorithm flowcharts.

\subsubsection{Computational Complexity}
The complexity of Alg.~\ref{alg:RRM} can be computed by adding the complexity of Alg.~\ref{alg:init_ua_ra} and the complexity of the iteration in Alg.~\ref{alg:RRM}.

The complexity of Alg.~\ref{alg:init_ua_ra} is $\mathcal{O}(I^2)$
as the maximum number of comparisons (Line 6 in Alg.~\ref{alg:init_ua_ra}) is $\frac{I(I+1)}{2}$.
The RA scheme in Alg.~3 in Appendix~\ref{Appendix:Optimal Resource Allocation} utilizes gradient descent with the complexity of $\mathcal{O}(\frac{I}{\epsilon})$ and the closed-form solution (69) in Appendix~\ref{Appendix:Optimal Power Control} takes $\mathcal{O}(1)$.
Then, the complexity of the Alg.~\ref{alg:RRM} is $\mathcal{O}(\frac{I^2}{\epsilon^2})$ since the iterations from Line 6 to Line 9 in Alg.~\ref{alg:RRM} takes $\mathcal{O}(\frac{1}{\epsilon})$.

\subsection{
Decision-Making Algorithms for Trajectory-Planning}
\label{sec:Decision-Making Algorithms for Trajectory-Planning}
A consecutive update of $\mathbf{q}^{(1)}, ... , \mathbf{q}^{(T)}$ provides a solution of the reformulated lower-bound problem $\bf{\mathdutchcal{P}2}$ ,
as function $f(\mathbf{q}^{(t)})$ now can be obtained by Alg.~\ref{alg:RRM}. 
The lower-bounded problem is considered a Markov decision process as illustrated in Fig.~\ref{fig:Markov representation}, where various optimization strategies exist.

We adopt a genetic algorithm (GA), limited depth-first search (DFS), and deep q-learning (DQN) as the main strategies.
A detailed algorithm design, characteristics, and summary of pros and cons can be arranged as follows:
\begin{itemize}
    \item 
    \textbf{GA-TP  \cite{weise2009-genetic} (Naive upper-bound)}: A canonical GA with elitism can find the global optimum after a sufficient number of generations and populations \cite{weise2009-genetic}.
    The gene is defined as a trajectory $\mathbf{Q}=[\mathbf{q}^{(t)}]^{t\in\mathcal{T}}$; and the fitness is defined as $\sum_{t\in\mathcal{T}} f(\mathbf{q}^{(t)})$.
    The genetic algorithm is configured as 10,000 generations, 10 elite counts, 50 populations, and 10\% mutation probability.
    Figure~\ref{fig:TP-RRM-visualization}b demonstrates the overall algorithm flows.
    In the elite selection, the fitness is computed for each gene.
    \item 
    \textbf{DFS \cite{CLRS_intro_to_algorithms}}: 
    DFS targets to find the best sub-trajectory $q^{(t)},...,q^{(t+n)}$ by the depth-first search and repeats the computation until DFS covers the entire time slots.
    As illustrated in Fig.~\ref{fig:TP-RRM-visualization}c, DFS considers all possible actions as a tree and finds the best sub-trajectory.
    Then, DFS can find the global optimal solution when $t=0$ and $n=T$.
    However, the practical depth $n$ is limited due to the limitations in the memory size and computing time \cite{CLRS_intro_to_algorithms}.
    We set the search depth $n$ as $\{1,3,5\}$.
    \item
    \textbf{DQN \cite{Mnih2015-HumanlevelCT}}:
    DQN is a well-known algorithm that can optimize a discrete MDP.
    To determine $\mathbf{q}^{(t)}$, we configured the state as a concatenation of $\mathbf{q}^{(t-1)}$, $[\mathbf{q}_i]_{i\in\mathcal{I}}$, $[s_i]_{i\in\mathcal{I}}$ $[T_i]_{i\in\mathcal{I}}$, $[\sum_{k=1}^{t-1}R_i^{(k)}]_{i\in\mathcal{I}}$; action space as $S(\mathbf{q}^{(t-1)})$; and reward $r(\cdot)$ as $r(\mathbf{q}^{(t)}) = f(\mathbf{q}^{(t)})$.
    Fig.~\ref{fig:TP-RRM-visualization}c visualizes states, rewards, and actions of DQN.
    Implementation details are provided in the Appendix~\ref{appendix:dqn}.
\end{itemize}

The complexity of the three TP schemes combined with the RRM scheme is computed by multiplying the complexity of the TP scheme and Alg.~\ref{alg:RRM}.
When $m$ generations and $n$ populations are given, the complexity of GA is $\mathcal{O}(mn\frac{I^2}{\epsilon^2})$.
For DFS, if the maximum number of reachable grid points $m$ with search depth $n$ is defined as $m=\max_{\mathbf{q}\in\mathcal{Q}} |S(\mathbf{q})|$, the complexity of DFS is $\mathcal{O}(mn\frac{I^2}{\epsilon^2})$.
The complexity of DQN is $\mathcal{O}(\frac{I^2}{\epsilon^2})$ as the feed-forward of the network takes $\mathcal{O}(1)$.
The key factor affecting computational complexity is the number of possible actions $|s(\mathbf{q})|$ in the MDP, as the TP algorithms are derived from the MDP formulation. In large-scale networks, the complexity of the TP algorithms can be reduced to a manageable level by adjusting the grid size $\Delta Q$ and the time slot length $\Delta T$.

\section{Simulation Results \label{sec:Simulation_Results}}

\begin{table}[htb]
    \centering
    \caption{Parameter Configurations}
    \begin{tabular}{cc}
    \toprule
    Parameter & Value \\ \cmidrule(r){1-1}\cmidrule{2-2}
    Map width $w$ (m)                                    & 600      \\ 
    Minimum altitude $h_{\text{min}}$ (m)                & 50       \\
    Maximum altitude $h_{\text{max}}$ (m)                & 200      \\
    Unit length of the grid map $\Delta Q$ (m)           & 40 \\
    Number of users $I$                                  & 10 to 80 \\
    Number of time slots $T$                              & 20     \\
    Unit length of the time slot $\Delta T$ (s)           & 3 \\
    Vehicle velocity $v$ (m/s)                           & 15       \\
    Carrier frequency (GHz) & 2 \\
    Bandwidth $B$ (MHz)                                  & \{2, 5, 10\}        \\ 
    Transmission power $P$ (dBm)                         & 23       \\
    Noise spectral density (dBm/Hz) & -173.8 \\
    Data rate constraint $r_i, \forall i \in \mathcal{I}$ (Mbps)    & 0 to 10  \\
    Environmental parameter $a$                          & 9.64     \\
    Environmental parameter $b$                          & 0.06     \\
    Excessive LoS pathloss $\eta_{\text{LoS}}$ (dB)  & 1        \\
    Excessive NLoS pathloss $\eta_{\text{NLoS}}$ (dB) & 40       \\ 
    Rician $K$-factor & 12 \\
    \bottomrule
    \end{tabular}
    \label{table:parameters}
\end{table}

In various performance metrics, we compare the proposed schemes with the other PF-maximizing schemes.
We first show that the proposed RRM converges to the global optimum of Problem \eqref{eq:lookahead_value}, then evaluate three control schemes (GA, DFS, and DQN).
Simulation parameters are chosen from existing works \cite{Matolak_Rician,A2GModel}, and 3GPP specifications \cite{3gpp.26.925}, which are listed in Table \ref{table:parameters}. 

In the experiments, the length of the request period $T_i,~\forall i \in \mathcal{I}$ and the initial request time $s_i,~\forall i \in \mathcal{I}$, follow the discrete uniform distributions, $\mathcal{U}[4,8]$ and $\mathcal{U}[0,T]$, respectively.

We have developed a TP simulator that implements the proposed method and several comparison schemes. The simulations are implemented under Python 3.9 on AMD Ryzen\textsuperscript{\texttrademark} 9 5950X processor.

To evaluate the performance of the RRM scheme Alg.~\ref{alg:RRM}, we additionally implement a genetic RRM (GA-RRM) and maximum SINR scheme (MAX SINR) that provide a solution $\mathbf{A}^{(t)}$, $\mathbf{B}^{(t)}$, and $\mathbf{P}^{(t)}$ in Problem \eqref{eq:lookahead_value}.
\begin{itemize}
    \item 
    \textbf{GA-RRM \cite{weise2009-genetic} (Naive upper-bound)}:
    The gene is defined as a concatenation of $\mathbf{A}^{(t)}$, $\mathbf{B}^{(t)}$, and $\mathbf{P}^{(t)}$; and the fitness is $f(\mathbf{q}^{(t)})$.
    The GA-RRM is configured as 10,000 generations, 10 elite counts, 100 populations, and 10\% mutation probability.
    \item
    \textbf{MAX SINR \cite{Jacob19-MAX_SINR_1, Asad23-MAX_SINR_2}}: 
    For the UA scheme, a user with the greatest SINR is associated with the UAV-BS to maximize the throughput.
    This UA scheme is prevalent in the sum-rate maximization problems \cite{Jacob19-MAX_SINR_1, Asad23-MAX_SINR_2}, providing a benchmark of the heuristic UA scheme.
    RA and PC schemes are the same as those of Alg.~\ref{alg:RRM}.
\end{itemize}

Then, we evaluate the proposed TP schemes in Sec.~\ref{sec:Decision-Making Algorithms for Trajectory-Planning} with the following comparison schemes:
\begin{itemize}
    \item
    \textbf{GA-ITER \cite{luenberger1984-alternating_optimization,Hao23-alternating_optimization} (Conventional iterative optimization)}
    Iterative optimization of the TP and RRM.
    The canonical GA and Alg.~\ref{alg:RRM} are used for TP and RRM, respectively.
    The GA parameters are identically configured as in GA-TP. 
    Properties are listed as follows:\\ 
    - \textit{Stationary RRM}: The RRM variables in GA-ITER are fixed when optimizing TP by GA. 
    However, the RRM in GA-TP changes for every gene. \\
    - \textit{Iterations}: Starting with a random trajectory, GA-ITER optimizes radio resources $\mathbf{A}$, $\mathbf{B}$, and $\mathbf{P}$ for a fixed trajectory.
    Then, the trajectory is optimized with the fixed $\mathbf{A}$, $\mathbf{B}$, and $\mathbf{P}$.
    Iteration stops either when there is no improvement in the objective \eqref{eq:MDP_objective} or when GA-ITER reaches 10 iterations\footnote{On average, GA-ITER converges in less than 5 iterations}.
    \item \textbf{OFDMA \cite{Zeng_OFDMA_relay_GLOBECOM, Zeng_OFDMA_relay_TWC}}:
    We implement the closest research\footnote{The energy consumption and relaying model in \cite{Zeng_OFDMA_relay_TWC} are not considered in our implementation.} that optimizes TP, UA, RA, and PC variables.
    The key differences are listed as follows:\\
    - \textit{The objective function}: The objective function is relaxed from the summation of a logarithm to the weighted sum-rate.
    For a fair comparison, the PF of all schemes are computed as a summation of logarithms \eqref{eq:p1_objective_function}. \\
    - \textit{QoS constraint}: The proposed scheme adopts the data rate as a QoS constraint, \cite{Zeng_OFDMA_relay_GLOBECOM} and \cite{Zeng_OFDMA_relay_TWC} adopt the SNR threshold.
    For a fair comparison, the SNR thresholds $\gamma^{\text{OFDMA}}_i,~i\in\mathcal{I}$ are chosen to satisfy $r_i = B\log_2(1+\gamma^{\text{OFDMA}}_i)$,
    where $\gamma^{\text{OFDMA}}_i$ implies the minimal SNR requirements for achieving data rate $r_i$.
    \\
    - \textit{Problem formulation}: The trajectory is obtained by sequentially optimizing the position for every time step, similar to DFS with tree depth $n=1$.
    \item \textbf{Circular-TP (Inspired by \cite{Wu_TWC_Circle, Ono_TWC_Circle})}: 
    While UA, RA, and PC are optimized by Alg. \ref{alg:RRM}, the UAV-BS position $\mathbf{q}^{(t)}$ is parametrized as 
    \begin{align}
        \mathbf{q}^{(t)} = 
        \begin{bmatrix}
            100\cos(\theta_t)+300 \\
            100\sin(\theta_t)+300 \\
            200
        \end{bmatrix},
    \end{align}
    where $\theta_t = \theta+\frac{v\Delta T}{r} t$ for randomly selected $\theta\in [0, 2\pi]$.
    Then, the variable $\theta$ determines the initial position, and the UAV-BS rotates the orbit with angular velocity $\frac{v\Delta T}{r}$.
    \item \textbf{Fixed-TP}: 
    For all $t\in\mathcal{T}$, the UAV-BS position is fixed at $\mathbf{q}^{(t)}=[300, 300, 200]^{\rm{T}}$.
    Same as Circular-TP, this scheme determines UA, RA, and PC by Alg. \ref{alg:RRM} for given position $\mathbf{q}^{(t)}$.
\end{itemize}

\subsection{Optimality of the Proposed RRM Scheme}

\begin{table}[ht]
\centering
\caption{
Regularized$^*$ $f(\mathbf{q}^{(t)})$(\%) and complexity of RRM schemes}
\label{tab:ua_comparison}

\begin{tabular}{cccccc}
\toprule[.5pt]

\multirow{2.25}{*}{\textbf{Algorithm}} & \multicolumn{4}{c}{\textbf{Number of users}} & 
\multirow{2.25}{*}{\textbf{Complexity}}{} \\
\cmidrule[.5pt](l{2pt}r{2pt}){2-5} &

\multicolumn{1}{c}{5} & \multicolumn{1}{c}{10} & \multicolumn{1}{c}{20} & \multicolumn{1}{c}{40} & \textbf{(Flops)} \\
\cmidrule(r{1pt}){1-1} \cmidrule(l{1pt}r{1pt}){2-5} \cmidrule(l{1pt}){6-6}
 
\rule{0pt}{8pt} \textbf{GA-RRM} & 100 & 100 & 100 & 100 & $\mathcal{O}(IG)^{**}$ \\ 
\rule{0pt}{8pt} 

\textbf{Alg.~\ref{alg:RRM} (Ours)} & 99.95 & 99.93 &  99.97 & 99.99 & $\mathcal{O}(I^2)$ \\
\rule{0pt}{8pt} 

\textbf{Max SINR} & 
73.54 & 55.19 & 43.91 & 37.78 & $\mathcal{O}(I)$ \\ 

\bottomrule[.5pt]
\end{tabular}
\vspace{3pt}

$*$ The PF values are regularized by those of GA-RRM. ~~~~~~~~~~~~~~~~~ \\
\hspace{-1pt}$**$ $G$ is the number of GA generations, where $I << G$.
~~~~~~~~~~~~~~
\end{table}

Once the proposed RRM scheme accomplishes global optimality, optimal decision-making for the proposed MDP formulation (as illustrated in Fig.~\ref{fig:Markov representation}) provides a global optimum of the lower-bound problem $\bf{\mathdutchcal{P}2}$.

Table~\ref{tab:ua_comparison} shows the regularized PF of the three RRM schemes to compute $f(\mathbf{q}^{(t)})$.
The number of users is assumed to be \{5, 10, 20, 40\}; QoS thresholds are configured as $r_i=5, i\in\mathcal{I}$ (Mbps); and initial data $R_i^{(0)}, i\in\mathcal{I}$ is randomly chosen from $\mathcal{U}[10, 30]$ (Mb).

\begin{figure}[htb]
    \centering
    \includegraphics[width=.9\columnwidth]{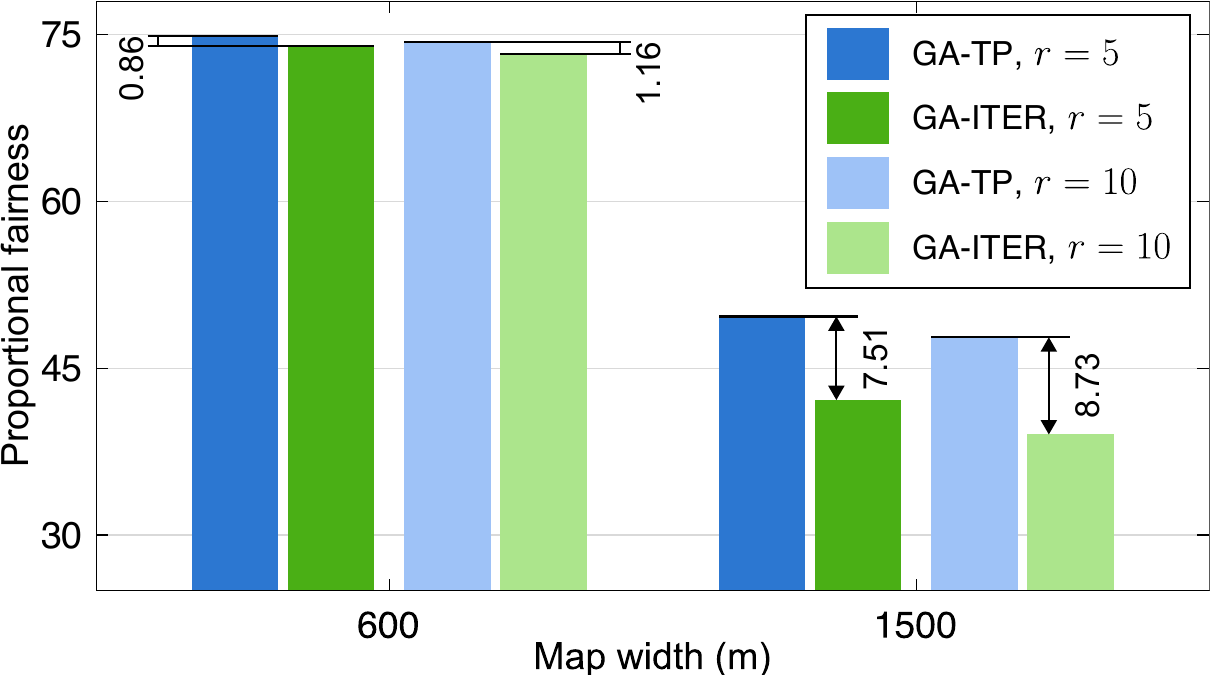}
    \caption{
    PF values of GA-ITER and GA-TP with bandwidth $B=10$ (MHz) and minimum data rate $r_i=r$ for $i\in\mathcal{I}$.}   \label{fig:ga_comparison}
\end{figure}

The proposed RRM scheme tightly achieves the global-optimal PF value, which is computed by GA-RRM.
Meanwhile, the PF values of the MAX SINR scheme decrease as the number of users increases.
From a PF standpoint, this is because providing service to multiple users, even at the cost of some sum-rate sacrifice, proves a superior strategy compared to prioritizing spectral efficiency.

\subsection{Measuring the Curse of Initialization}
Figure~\ref{fig:ga_comparison} illustrates PF differences in GA-TP and GA-ITER.
GA-TP and GA-ITER share the same GA parameters for TP; and Alg.~\ref{alg:RRM} for RRM.
However, the PF gap increases when the map width and minimum data requirements increase.
These factors affect both the user pathloss and candidate users whose QoS might be satisfied, 
so consequently determining user association within a given trajectory.
Therefore, GA-ITER is more likely to converge to the local optimum as the environment becomes rigorous, which leads to an increase in the PF gap.

\begin{figure*}[htb]
    \centering
    \null\hfill
    \subfloat[ $\mathbf{q}^{(0)}=\text{[}450, 450, 120\text{]}^{\rm{T}}$ \label{subfig:initial_position_2}]{\includegraphics[width = .3\textwidth, valign=c]{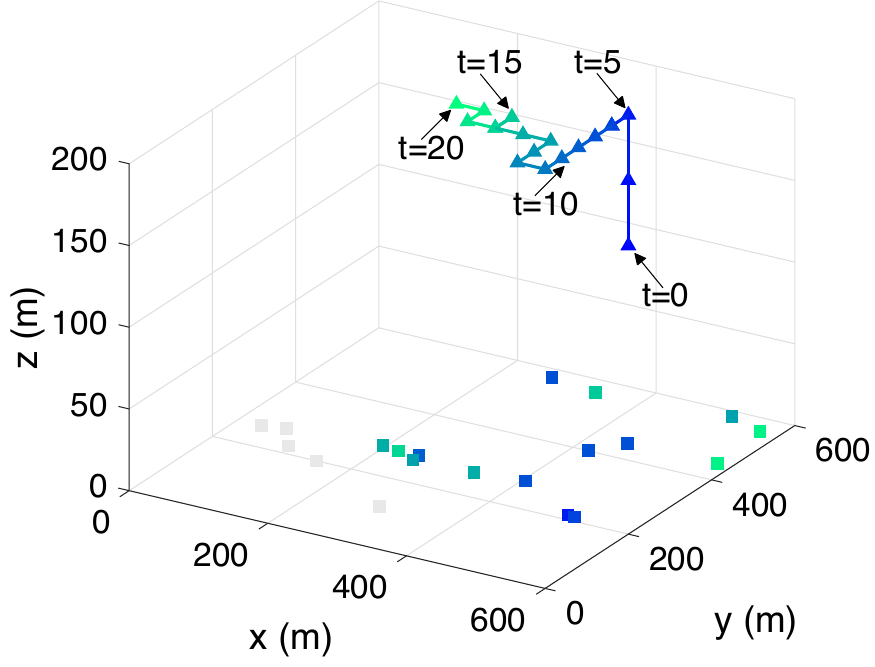}}
    \hfill
    \subfloat[$\mathbf{q}^{(0)}=\text{[}0, 400, 160\text{]}^{\rm{T}}$ \label{subfig:initial_position_3}]{\includegraphics[width = .3\textwidth, valign=c]{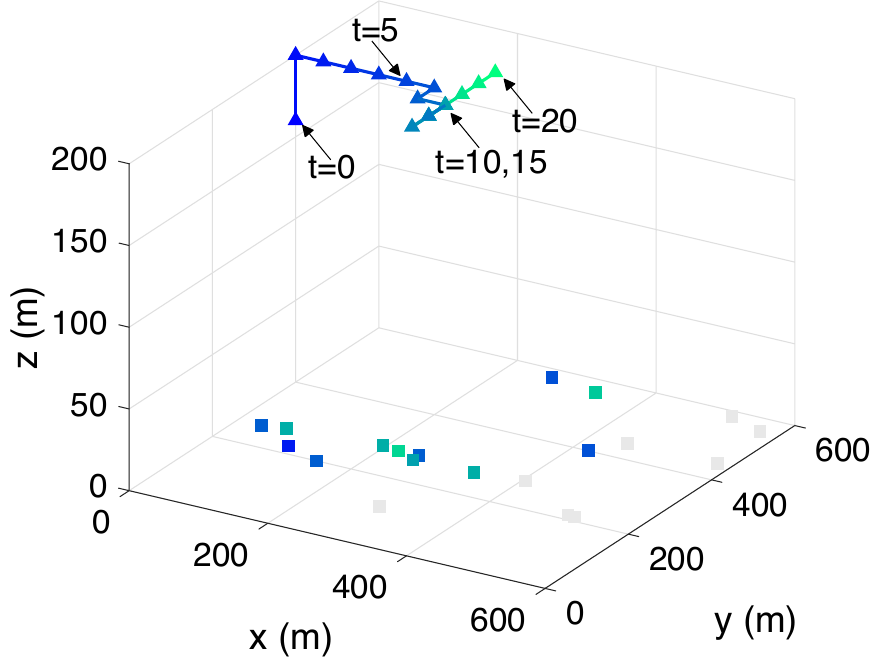}}
    \hfill
    \subfloat[$\mathbf{q}^{(0)}=\text{[}0, 80, 160\text{]}^{\rm{T}}$\label{subfig:initial_position_1}]{\includegraphics[width = .3\textwidth, valign=c]{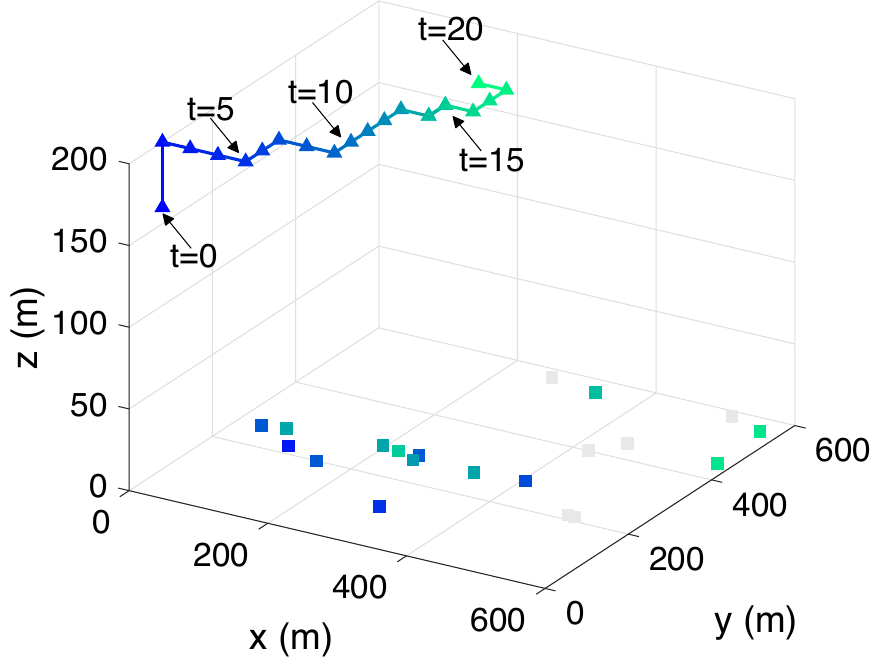}}
    \hfill
    \subfloat{\includegraphics[width = .07\textwidth, valign=c]{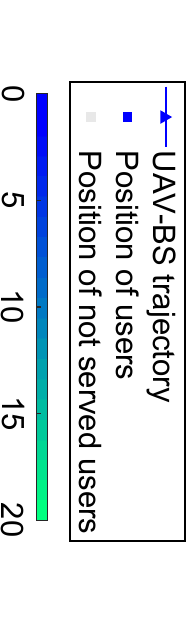}}
    \hfill\null
    \caption{
    Trajectories obtained from DQN with the same environment configurations, except for the initial position $\mathbf{q}^{(0)}$.}
    \label{fig:initial_position}
\end{figure*}

\subsection{
Adaptive Trajectory-Planning of MDP formulation
}

Figure~\ref{fig:initial_position} demonstrates that the proposed MDP formulation effectively resolves the dependency on the initial choice.
In the early motivation presented in Fig.~\ref{fig:toy_example}, we have argued that the initial position choice fetters subsequent variable optimization, thereby leading to a deterioration in network utility.
However, Fig.~\ref{fig:initial_position} illustrates that the MDP formulation enables the UAV-BS to make an adaptive trajectory regardless of the initial position.

\begin{figure*}[ht]
    \centering
    \null\hfill
    \subfloat[Bandwidth $B = 2 $ (MHz) \label{subfig:pf-rate-bw2}]{\includegraphics[width = .3\textwidth, valign=t]{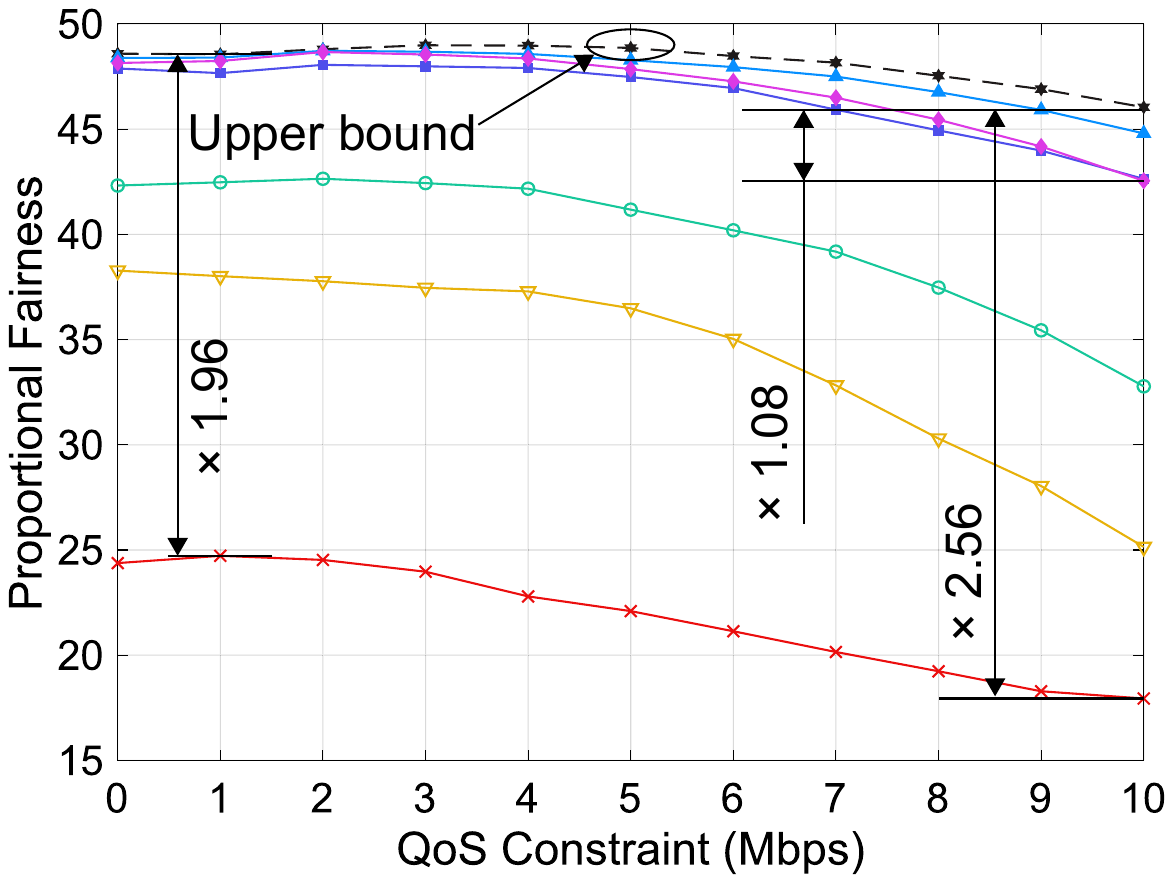}}
    \hfill
    \subfloat[Bandwidth $B = 5 $ (MHz) \label{subfig:pf-rate-bw5}]{\includegraphics[width = .3\textwidth, valign=t]{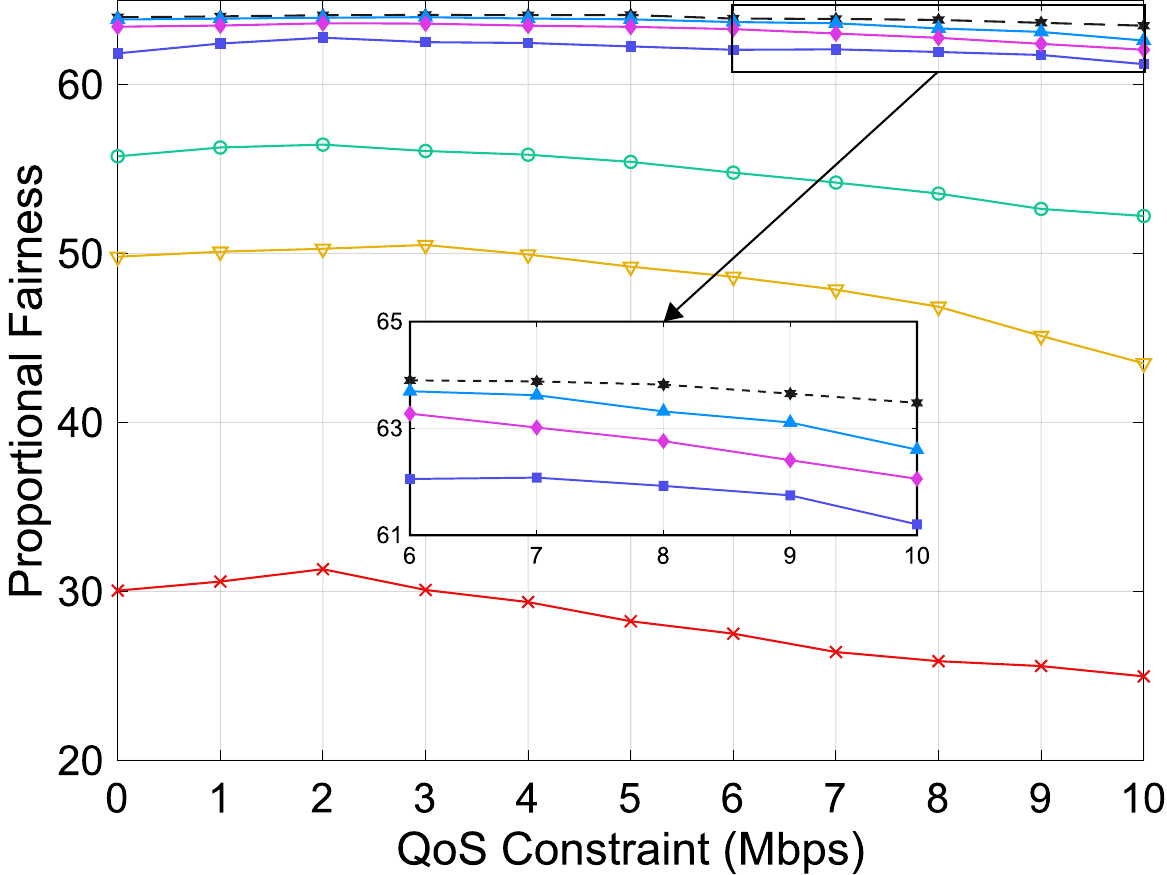}}
    \hfill
    \subfloat[Bandwidth $B = 10 $ (MHz) \label{subfig:pf-rate-bw10}]{\includegraphics[width = .3\textwidth, valign=t]{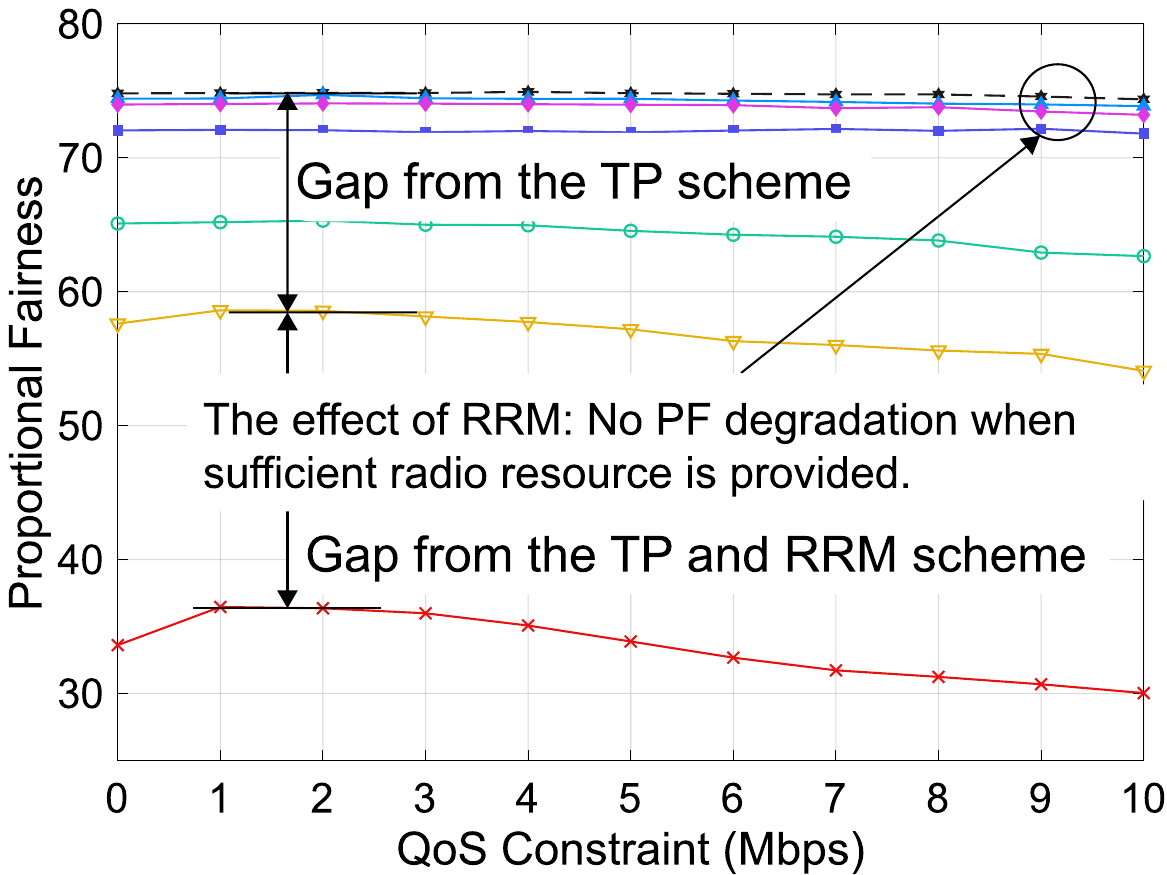}}
    \hfill
    \subfloat{\includegraphics[width = .09\textwidth, valign=t]{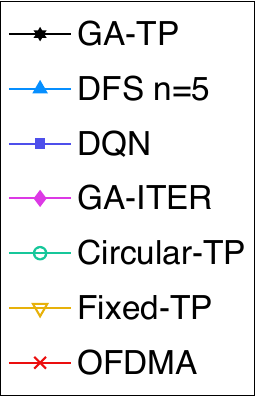}}
    \hfill\null
    \caption{
    Proportional fairness for $20$ users with various QoS constraints.
    }
    \label{fig:pf-rate}
\end{figure*}
\subsection{Proportional Fairness for Various QoS Constraints \label{Sec:PF-rate}}
Figure~\ref{subfig:pf-rate-bw2} depicts the PF for 20 users under various QoS constraints when the bandwidth is 2 MHz.
The figure shows that the three proposed TP schemes achieve more than $90\%$ higher PF value than the OFDMA scheme in the worst case.
This is because the OFDMA scheme maximizes the PF in the format of the weighted sum-rate.
Then, the OFDMA scheme's objective becomes linear in terms of the users' sum-rate, providing all resources to the most effective user could be an optimal strategy.
However, the strategy decreases the fairness of the time-critical mobile network as the users' request periods expire while a few users are selectively serviced.

We remark that the proposed schemes show a smaller decline in the PF value than the comparison schemes, even at the stringent QoS constraints. 
Compared with the greatest PF value of each scheme in Fig.~\ref{subfig:pf-rate-bw2}, the PF values at QoS constraints of 10 Mbps decrease 8\% for GA-TP and DFS; 11\% for DQN; 14\% for GA-ITER 23\% for Circular-TP; 34\% for Fixed-TP; and 27\% for the OFDMA scheme, respectively.

Figs.~\ref{subfig:pf-rate-bw5} and~\ref{subfig:pf-rate-bw10} depict the PF values with 5 and 10 MHz bandwidth, respectively.
The proposed schemes show consistent PF values regardless of the QoS constraints, while the PF values of Circular-TP and Fixed-TP slightly decrease.
This implies that the proposed schemes benefit from the high SNR channels because Circular-TP and Fixed-TP share the same RRM scheme with GA-TP. 

\begin{figure*}[t]
    \centering
    \null\hfill
    \subfloat[Bandwidth $B = 2 $ (MHz) \label{subfig:coverage-rate-bw2}]{\includegraphics[width = .30\textwidth, valign=t]{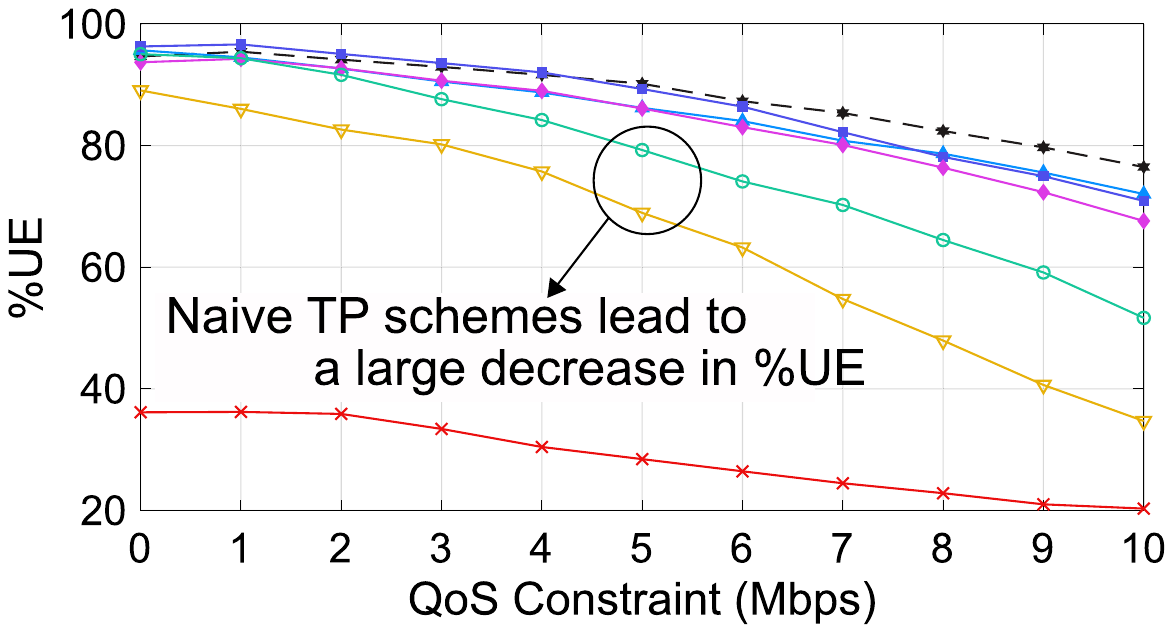}}
    \hfill
    \subfloat[Bandwidth $B = 5 $ (MHz) \label{subfig:coverage-rate-bw5}]{\includegraphics[width = .30\textwidth, valign=t]{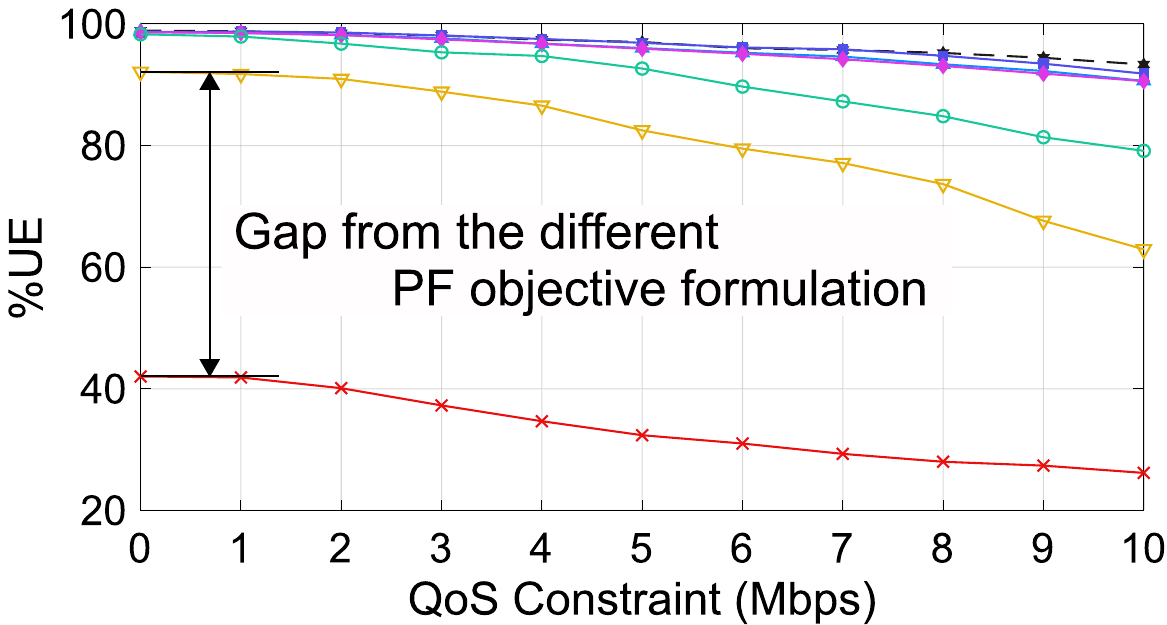}}
    \hfill
    \subfloat[Bandwidth $B = 10 $ (MHz) \label{subfig:coverage-rate-bw10}]{\includegraphics[width = .30\textwidth, valign=t]{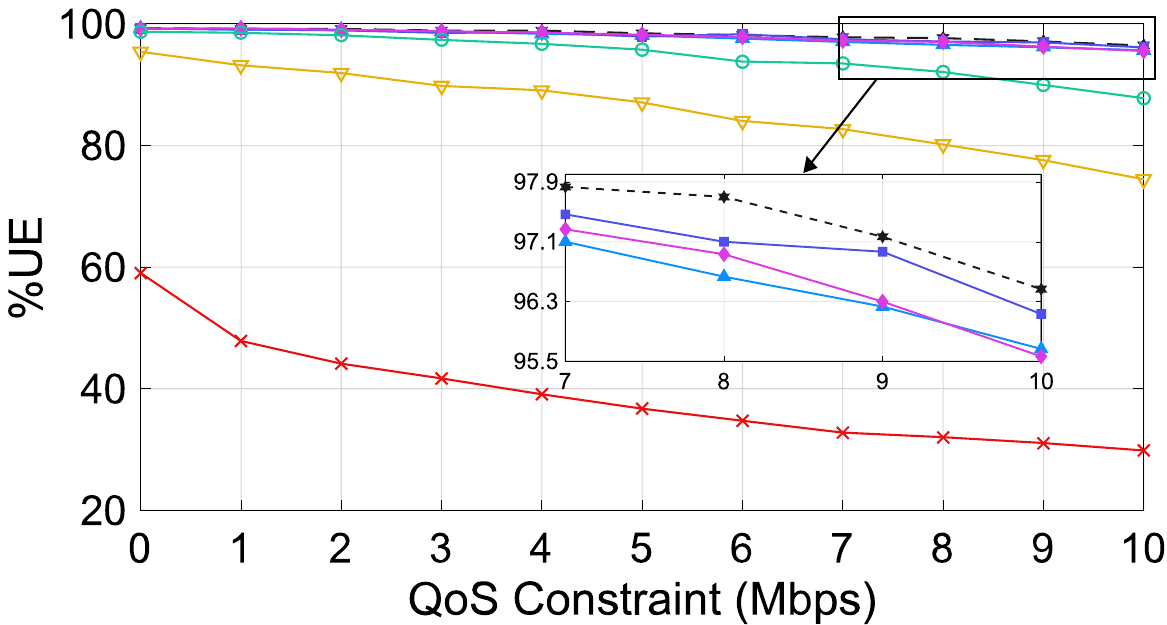}}
    \hfill
    \subfloat{\includegraphics[width = .09\textwidth, valign=t]{Figures/Results/legend-eps-converted-to.pdf}}
    \hfill\null
    \caption{
    Percentage of served users with various QoS constraints.
    The total number of users is configured to be 20.
    }
    \label{fig:coverage-rate}
\end{figure*}

\begin{figure*}[htb]
    \centering
    \hfill
    \subfloat[GA-TP \label{subfig:visualize-GA-TP}]{\includegraphics[width = .25\textwidth]{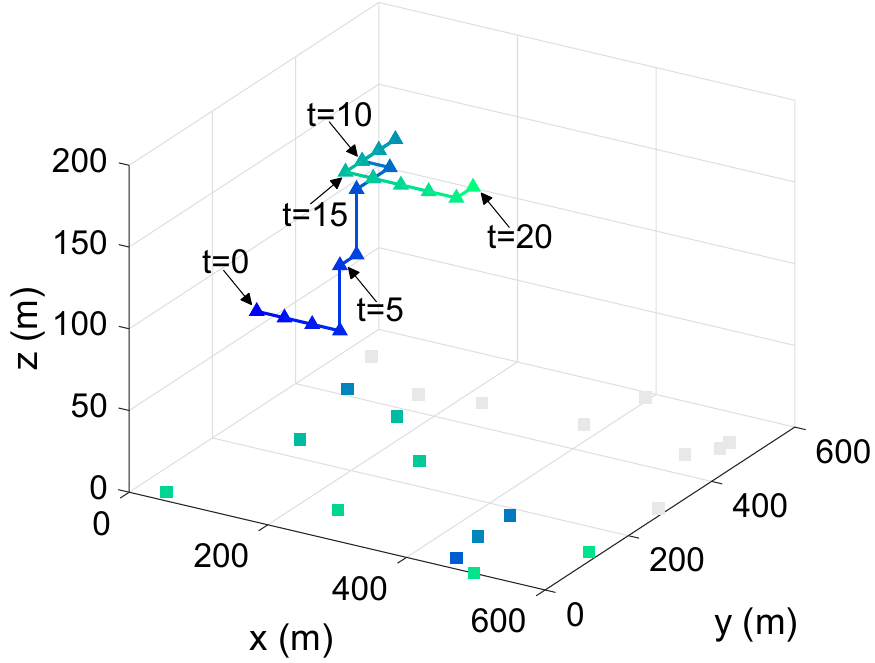}}
    \subfloat[DQN \label{subfig:visualize-DQN}]{\includegraphics[width = .25\textwidth]{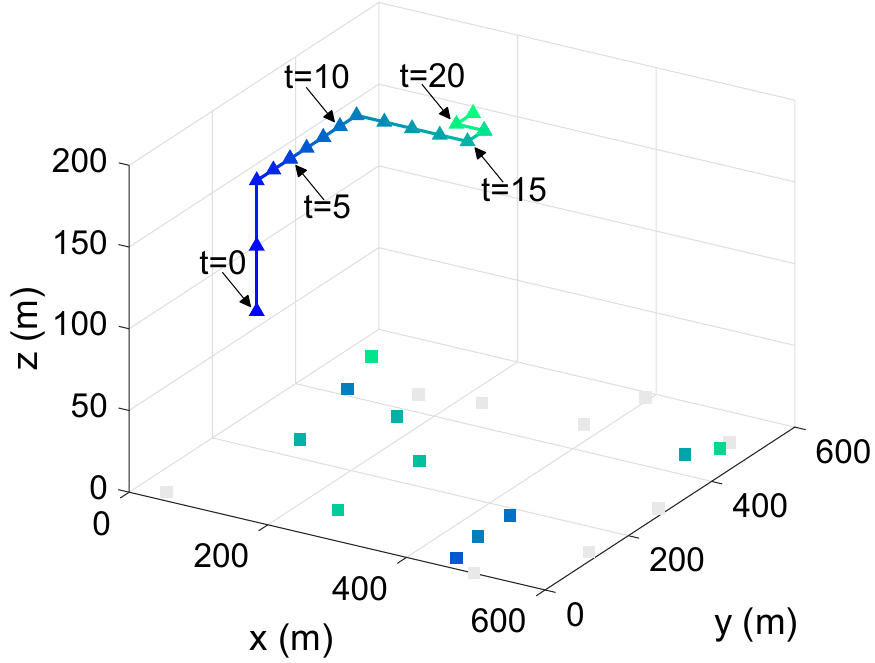}}
    \subfloat[GA-ITER \label{subfig:visualize-GA-ITER}]{\includegraphics[width = .25\textwidth]{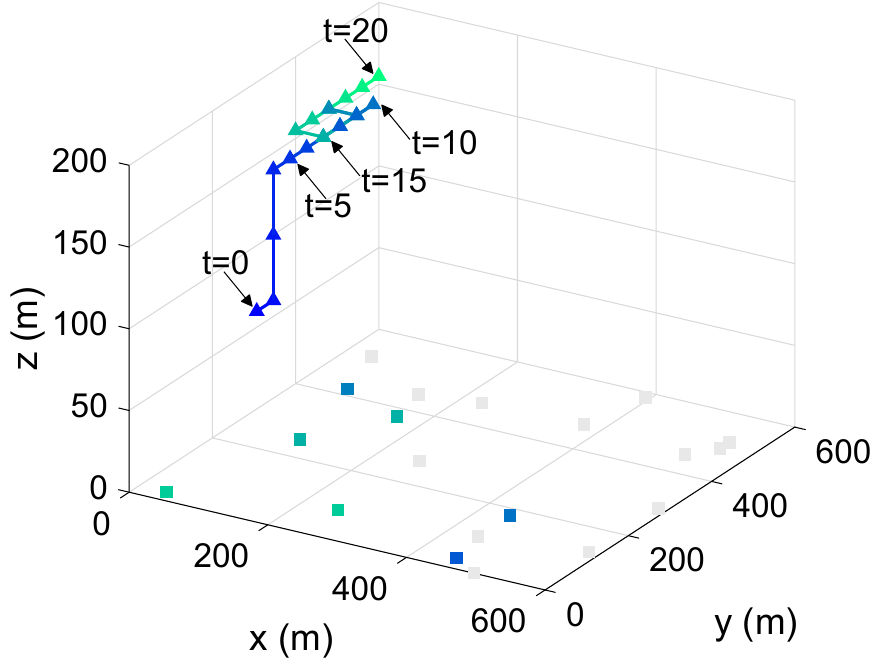}}
    \subfloat[OFDMA \label{subfig:visualize-OFDMA}]{\includegraphics[width = .25\textwidth]{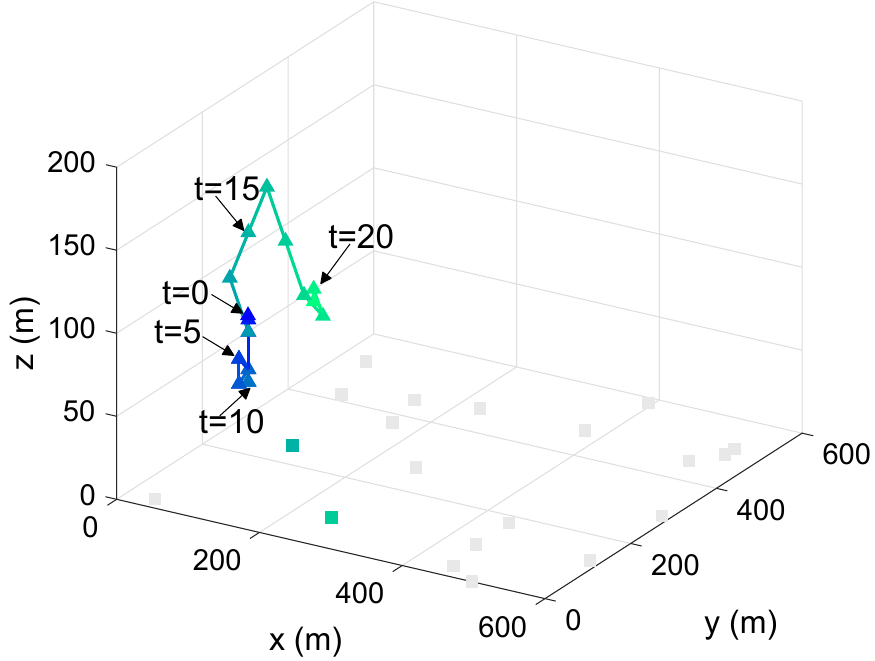}}
    \hfill \vspace{-6pt}
    \medskip
    \vspace{-6pt}
    \subfloat[DFS $n=1$ \label{subfig:visualize-n1}]{\includegraphics[width = .25\textwidth]{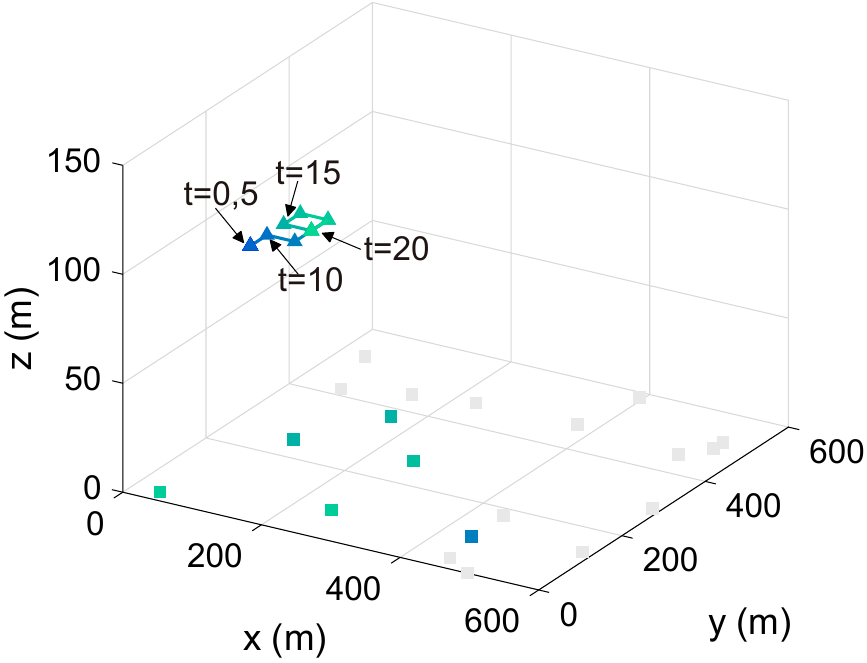}}
    \subfloat[DFS $n=3$ \label{subfig:visualize-n3}]{\includegraphics[width = .25\textwidth]{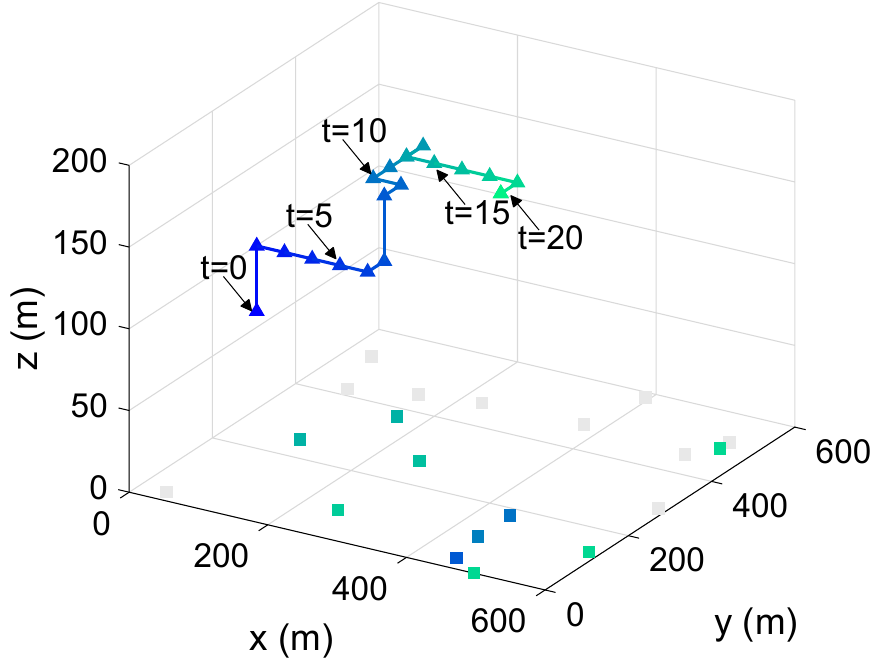}}
    \subfloat[DFS $n=5$ \label{subfig:visualize-n5}]{\includegraphics[width = .25\textwidth]{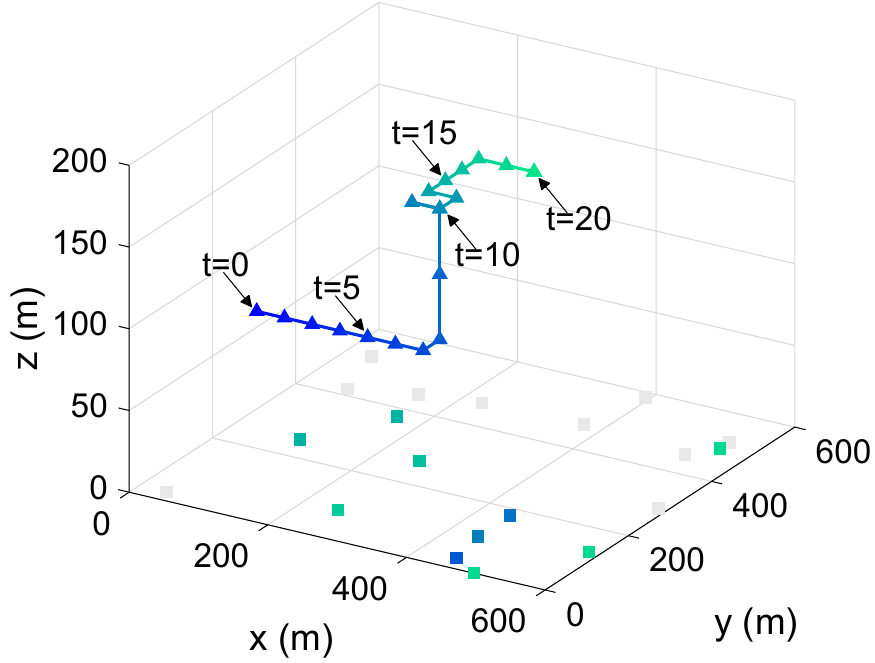}}
    \subfloat{\includegraphics[width = .25\textwidth]{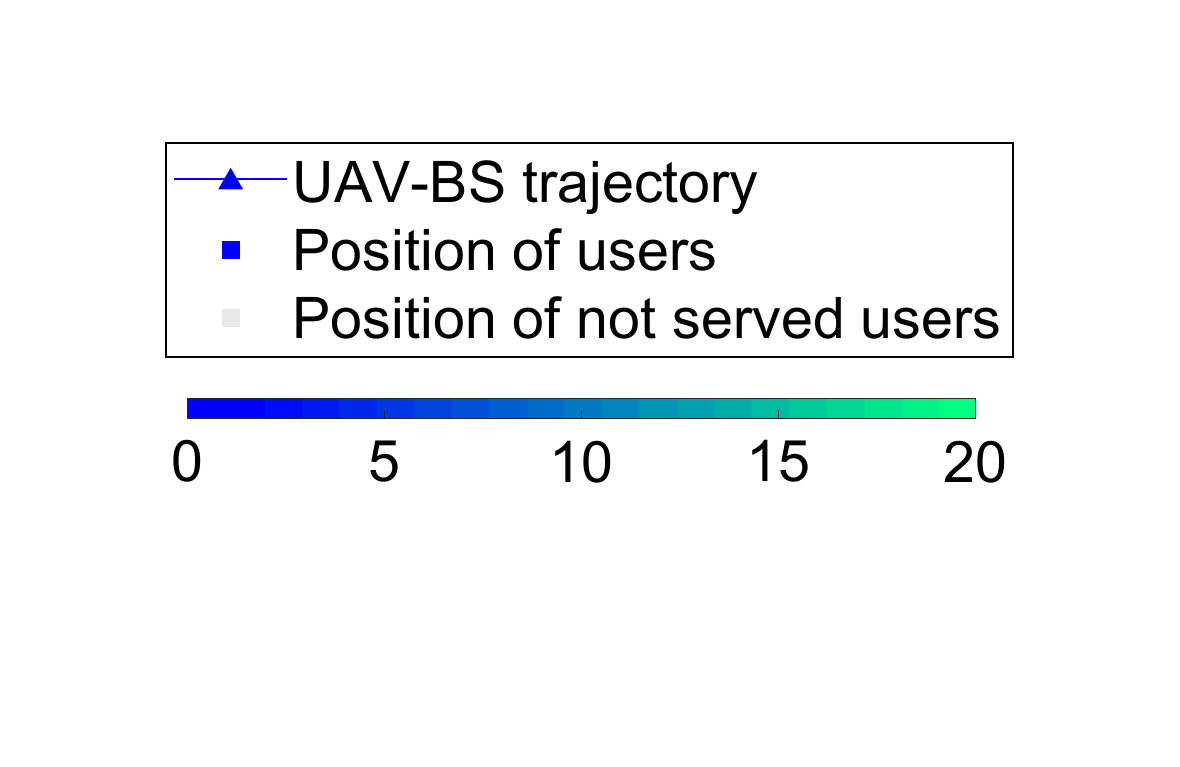}}
    \caption{
    Visualized trajectories for various TP schemes.
    The colors of the UAV-BS trajectory represent the index of time slots, 
    and the colors in the position of users represent the first time slot where the user receives service.
    The trajectory of the OFDMA scheme is not latticed as we preserve the system models of \cite{Zeng_OFDMA_relay_TWC, Zeng_OFDMA_relay_GLOBECOM}.
    }
    \label{fig:visualize}
\end{figure*}

\subsection{Percentage of Served Users for Various QoS Constraints}
Figure~\ref{fig:coverage-rate} illustrates the percentage of served users (\%UE) for various QoS constraints.
The percentage is calculated by dividing the count of users served at least once by the total number of users.
The overall tendency shows that the \%UE decreases as the QoS constraint increases.

The GA-TP scheme serves at least \{56, 57, 40\}\%p more users even in the worst case than the OFDMA scheme for bandwidth $\{2, 5, 10\}$ MHz, respectively.
This gap is mainly derived from the different formulations of the objective function.
As mentioned in the previous section \ref{Sec:PF-rate}, serving a few users could be an optimal UA policy in the weight sum-rate formulation.
However, the objective of the proposed problem is a summation of logarithms.
Then, concentrating all resources on a single user is not optimal because the growth of the logarithmic function diminishes with the increasing user sum-rate.
Therefore, multiple users are highly likely associated with the UAV-BS at a single time slot in the proposed scheme.

For Circular-TP and Fixed-TP, the gap from GA-TP increases as users' QoS constraints increase.
The proposed scheme shows a small gap in unconstrained cases,
but the gap increases up to 20\%p for Circular-TP and 37\%p for Fixed-TP as QoS constraints are configured to be $r_i=10,~\forall i\in\mathcal{I}$.

As depicted in Figs.~\ref{subfig:coverage-rate-bw5} and~\ref{subfig:coverage-rate-bw10}, the proposed schemes gradually converge to serving 100\% users regardless of the lookahead variable $n$ when the available bandwidth increases.
However, the OFDMA scheme shows little enhancements due to the winner-take-all tendency in the user association.
\subsection{Comparisons of the Trajectory and User Association}
Figure~\ref{fig:visualize} demonstrates the trajectories and user associations of various schemes.
We assume that $20$ users exist in the service area with QoS constraints $r_i=10$ Mbps for $i\in\mathcal{I}$ with the 2 MHz bandwidth.

As we argue in the previous two sections, we can observe that the OFDMA scheme takes the winner-take-all strategy, serving only the adjacent users adjacent to the UAV-BS's position in Fig.~\ref{subfig:visualize-OFDMA}.

In Figs.~\ref{subfig:visualize-GA-TP},~\ref{subfig:visualize-DQN}, and~\ref{subfig:visualize-n5}, we can observe a similar UA and TP tendency as the schemes closely converge to the optimum. 
The number of served users for DFS accordingly increases as $n$ increases.
This is because UAV-BS can be pre-positioned near users on demand, taking into account the user's future requirements.
The difference in trajectories accumulates along the service timeline, so
there is a big gap in the number of served users between $n=1$ and $n=5$ as the service time of the UAV-BS increases.

\begin{figure*}[tb]
    \centering
    \null\hfill
    \subfloat[Bandwidth $B = 2 $ (MHz) \label{subfig:pf-user-bw2}]{\includegraphics[width = .3\textwidth, valign=t]{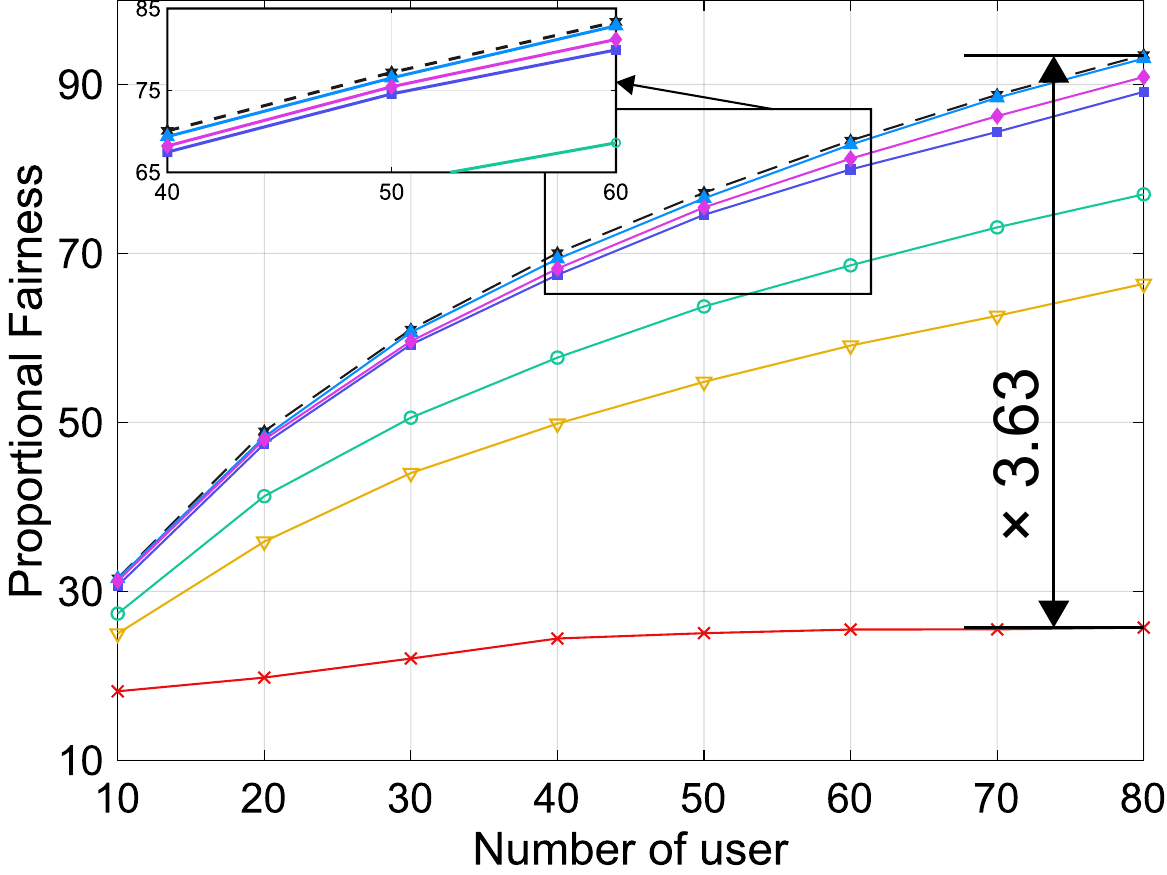}}
    \hfill
    \subfloat[Bandwidth $B = 5 $ (MHz) \label{subfig:pf-user-bw5}]{\includegraphics[width = .3\textwidth, valign=t]{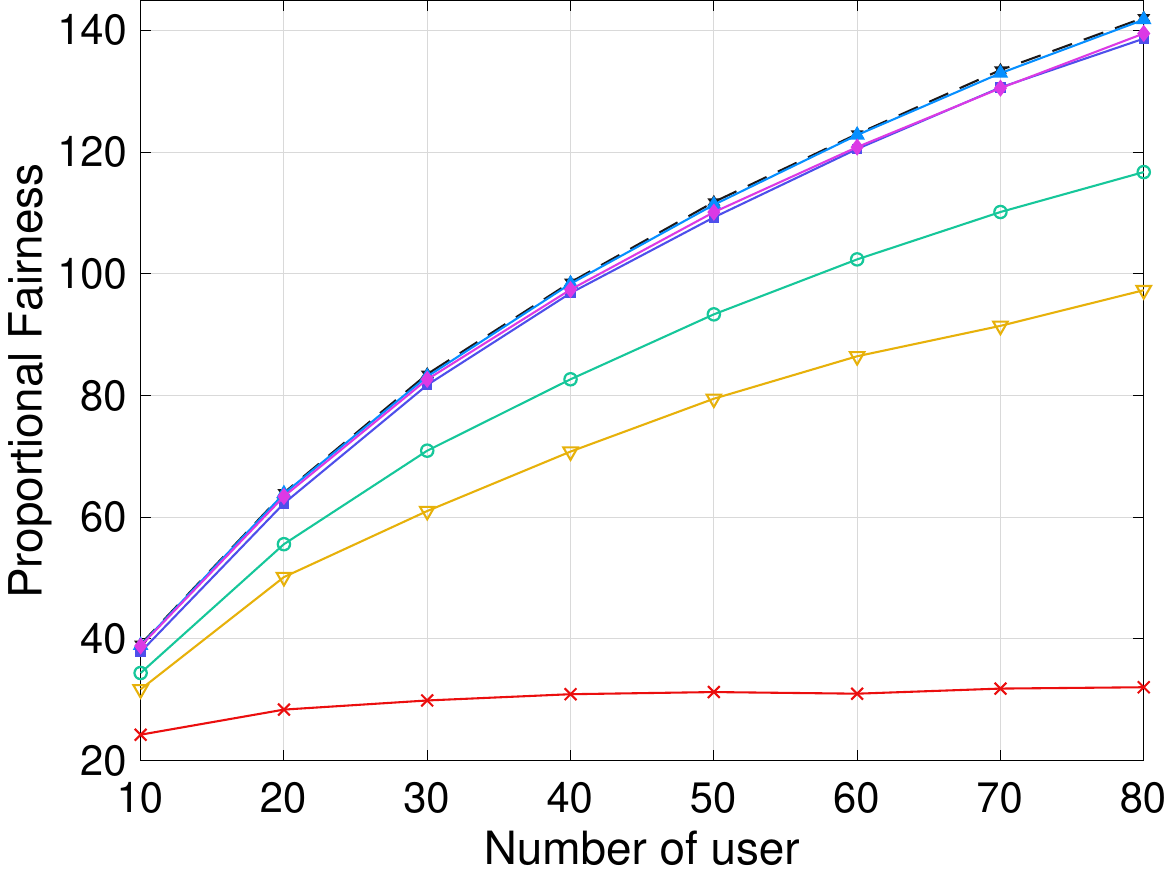}}
    \hfill
    \subfloat[Bandwidth $B = 10 $ (MHz) \label{subfig:pf-user-bw10}]{\includegraphics[width = .3\textwidth, valign=t]{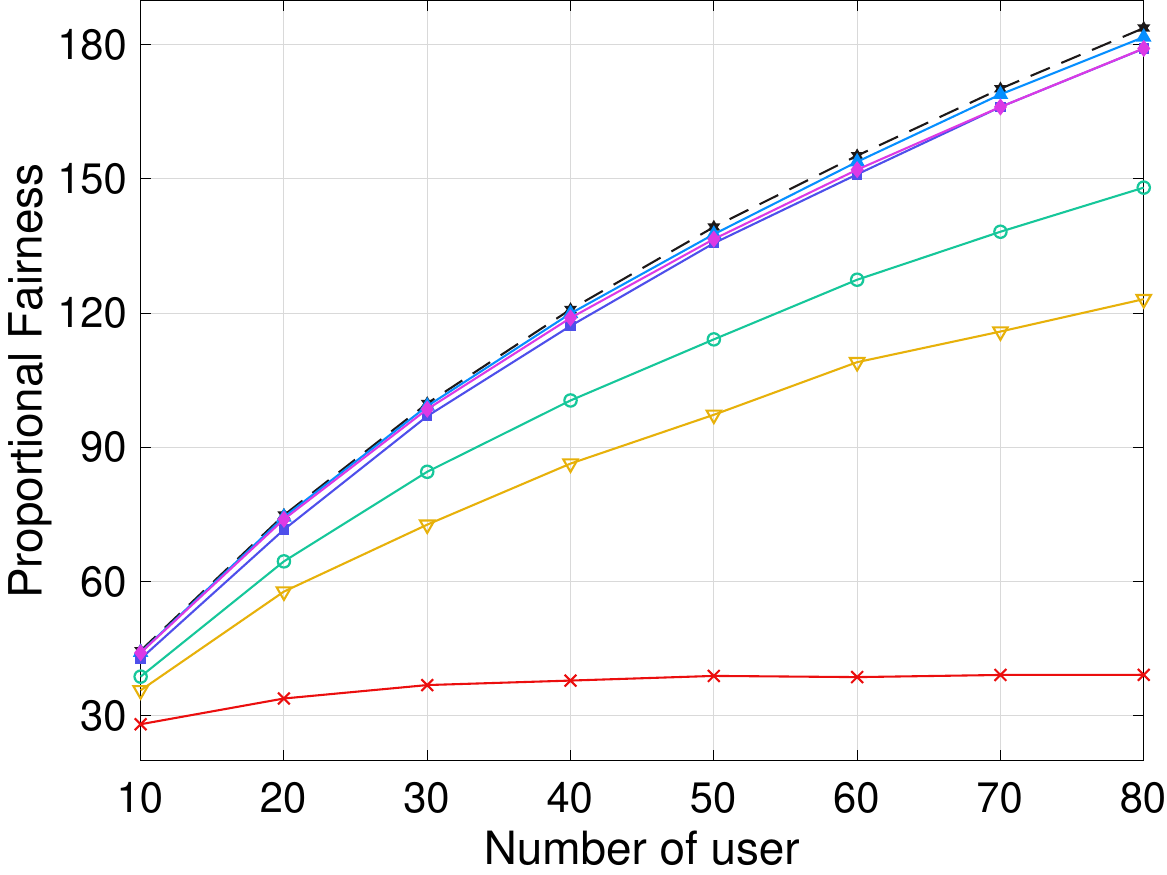}}
    \hfill
    \subfloat{\includegraphics[width = .09\textwidth, valign=t]{Figures/Results/legend-eps-converted-to.pdf}}
    \hfill\null
    \caption{
    Proportional fairness with a various number of users. 
    The rate constraints $r_i,~\forall i\in\mathcal{I}$ are configured as $r_i=5,~\forall i\in \mathcal{I}.$
    }
    \label{fig:pf-user}
\end{figure*}

\begin{figure*}[tb]
    \centering
    \null\hfill
    \subfloat[Bandwidth $B = 2 $ (MHz) \label{subfig:coverage-user-bw2}]{\includegraphics[width = .30\textwidth, valign=t]{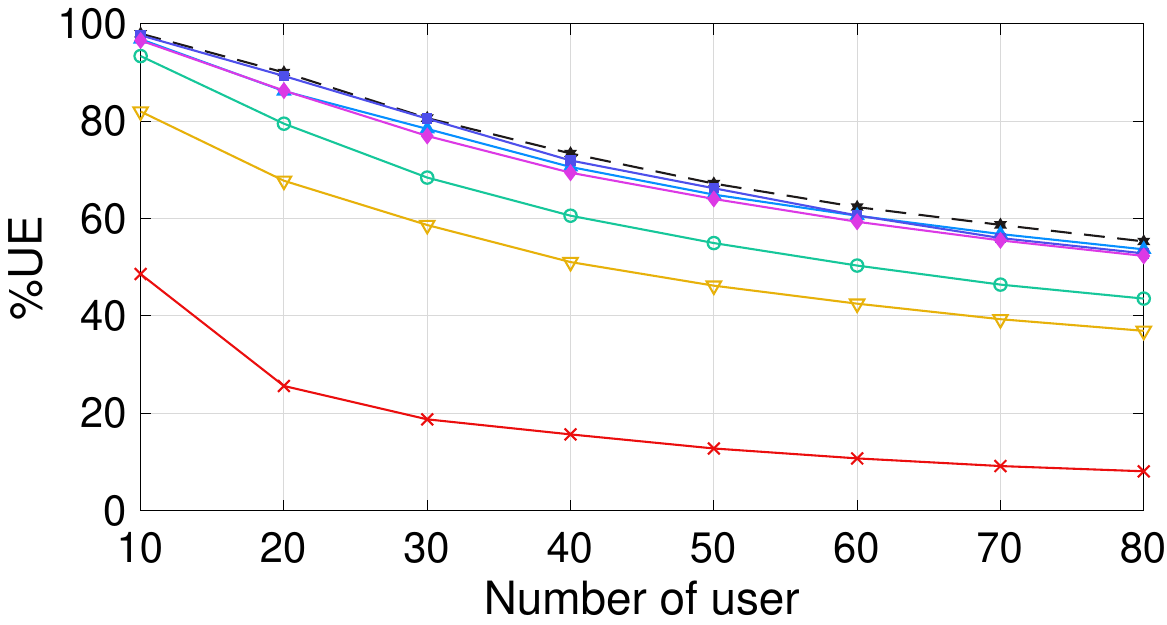}}
    \hfill
    \subfloat[Bandwidth $B = 5 $ (MHz) \label{subfig:coverage-user-bw5}]{\includegraphics[width = .30\textwidth, valign=t]{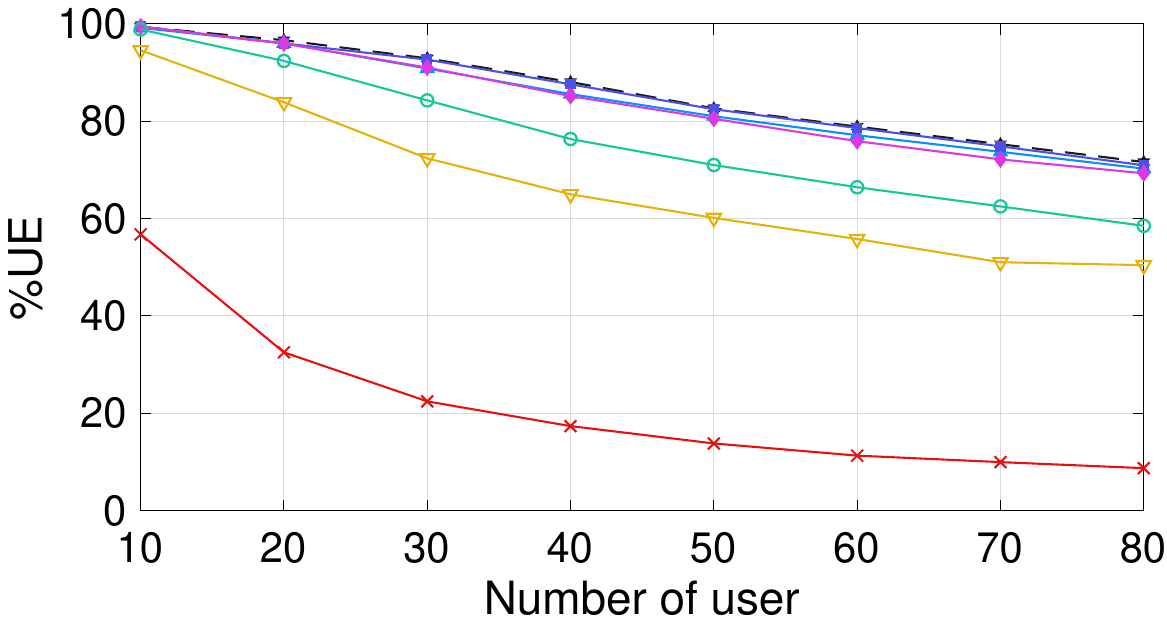}}
    \hfill
    \subfloat[Bandwidth $B = 10 $ (MHz) \label{subfig:coverage-user-bw10}]{\includegraphics[width = .30\textwidth, valign=t]{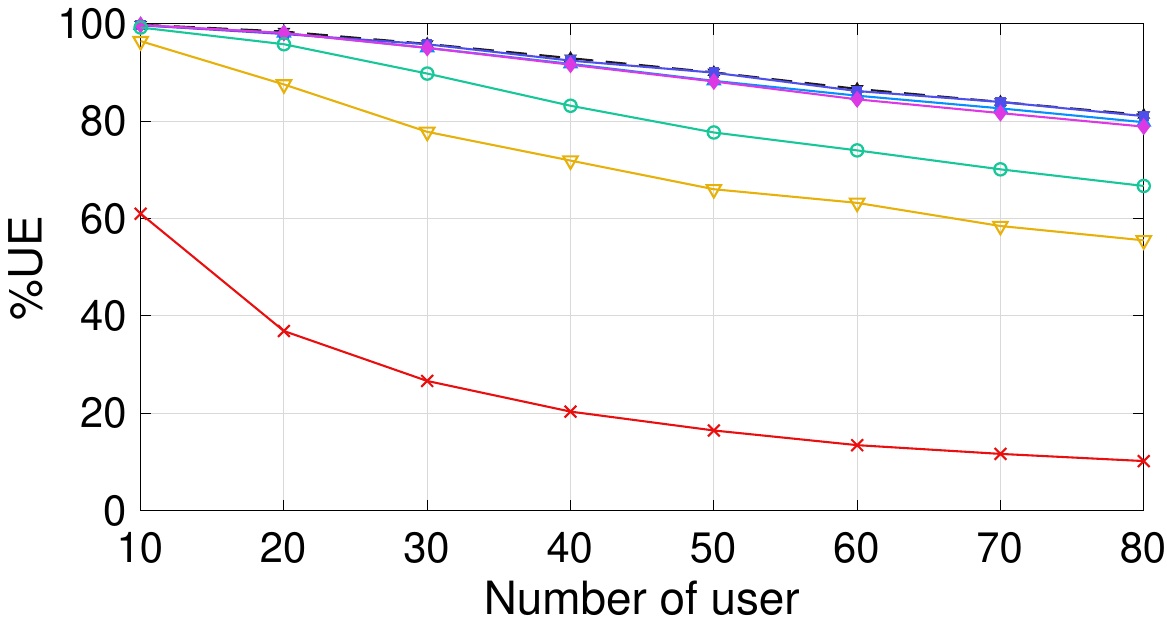}}
    \hfill
    \subfloat{\includegraphics[width = .088\textwidth, valign=t]{Figures/Results/legend-eps-converted-to.pdf}}
    \hfill\null
    \caption{
    Percentage of served users with a various number of users. }
    \label{fig:coverage-user}
\end{figure*}
\subsection{Proportional Fairness for Various Number of Users}
Figure~\ref{fig:pf-user} presents the PF for $\{10, 20, ..., 80\}$ users with  $\{2, 5, 10\}$ MHz bandwidth.
The QoS constraints are assumed to be $r_i=5$ Mbps for all $i\in\mathcal{I}$.

The PF of the $\{$GA-TP, DFS, DQN, GA-ITER$\}$ schemes increases up to $\{296, 295, 290, 290\}$\% as the number of users increases (Fig.~\ref{subfig:pf-user-bw2}).
Also, Circular-TP and Fixed-TP have similar increases, but they show 18\% and 40\% lower PF than DFS, respectively.
This implies that the proposed schemes benefit from the high SNR channels obtained from the TP scheme
because Circular-TP and Fixed-TP share the same RRM scheme with the proposed schemes.
Meanwhile, the PF of the OFDMA scheme shows a 41\% increase.

As illustrated in Fig.~\ref{subfig:pf-user-bw5}, the PF for 5 MHz bandwidth shows a similar trend with that for 2 MHz bandwidth (Fig.~\ref{subfig:pf-user-bw2}).
However, the PF gap between the proposed schemes and the OFDMA scheme increases as the available bandwidth increases.
The GA-TP and DFS schemes show around 60\% higher PF value at 10 users and 363\% higher PF value at 80 users than the OFDMA scheme.
Similarly, when the UAV-BS has 10 MHz bandwidth, the PF gap between the DFS and OFDMA scheme increases from 57\% for 10 users to 264\% for 80 users in Fig.~\ref{subfig:pf-user-bw10}.

\subsection{Percentage of Served Users for the Various Number of Users}
Figure~\ref{fig:coverage-user} depicts \%UE for $\{10, 20, ..., 80\}$ users with  $\{2, 5, 10\}$ MHz bandwidth.
The QoS constraints are configured to be $r_i=5$ Mbps for all $i\in \mathcal{I}$.

We can compute the average number of served users from Fig.~\ref{fig:coverage-user}.
In Fig.~\ref{subfig:coverage-user-bw2}, the average number of served users increases from 9.8 to 44.24 for GA-TP; from 9.68 to 42.93 for DFS; from 9.76 to 42.26 for DQN; and from 9.66 to 41.85 for GA-ITER.
Meanwhile, the average number of served users for the OFDMA scheme increases from 4.85 to 6.43.
This implies that the increase in Fig.~\ref{subfig:pf-user-bw2} of the OFDMA scheme is mainly caused by the channel enhancement from the high user density, not by the increase in the number of served users.

In Fig.~\ref{subfig:coverage-user-bw5}, we note that the overall percentage of served users increases when the UAV-BS utilizes 5 MHz bandwidth.
\{GA-TP, DFS, DQN\} achieve about \{16, 17, 18\}\%p higher \%UE for 5 MHz bandwidth than for 2 MHz, 
but the OFDMA scheme shows almost no difference regardless of the size of available bandwidth.
Figure~\ref{subfig:coverage-user-bw10} shows that the percentage of served users gradually approaches 100\%, having a similar tendency with Figs.~\ref{subfig:coverage-user-bw2} and \ref{subfig:coverage-user-bw5}.

\begin{figure}[tb]
    \centering
    \includegraphics[width=\columnwidth]{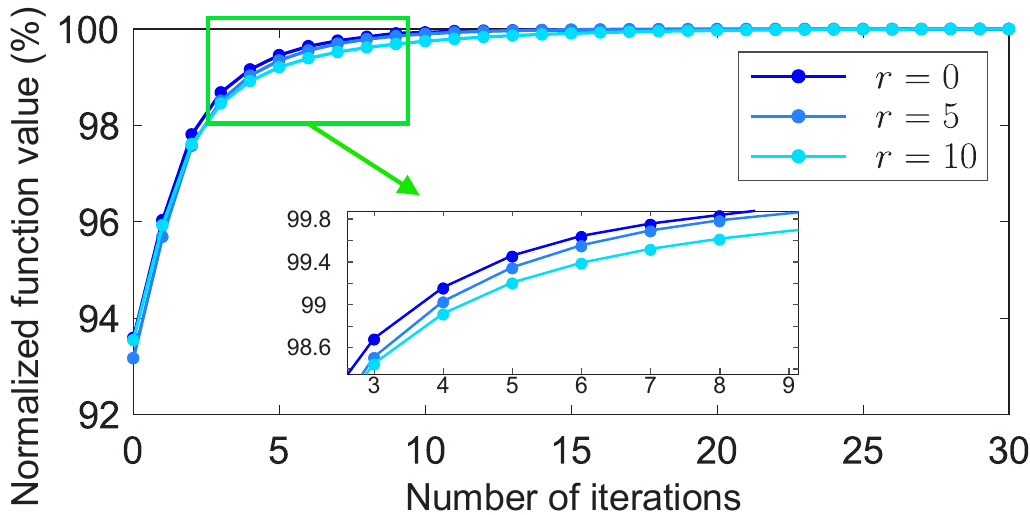}
    \caption{The output value of Alg.~\ref{alg:RRM}, normalized by the convergence value.
    The QoS constraints are configured as $r_i=r, \forall i\in \mathcal{I}$.
    }
    \label{fig:convergence}
\end{figure}

\subsection{Convergence and Computational Complexity Analysis}
Figure~\ref{fig:convergence} depicts the regularized PF for each iteration in the RRM scheme (Alg.~\ref{alg:RRM}).
Zero iteration implies that the UA and RA variables are only optimized by the initial UA and RA scheme (Alg.~\ref{alg:init_ua_ra}) without implementing Alg.~\ref{alg:RRM}.
Even at the zero iteration, Alg.~\ref{alg:init_ua_ra} achieves 93\% of the regularized PF value.
Then, the PF accomplishes 99\% and 99.9\% after 5 and 10 iterations, respectively.

\section{Conclusion}
\label{sec:Conclusion}
This paper proposes a separation of control and RRM in the time-critical aerial network.
We find a critical drawback of conventional joint iterative optimizations, where the initial trajectory choice fetters the subsequent variable choices.
However, separating TP and RRM in the practical scenario is challenging because the user requirements and fairness tightly bind network variables.
We address the challenge by decomposing the proposed problem into the sum of serial sub-problems.
Then, we cleverly transform the sub-problems into the MDP formulation where the TP and RRM variables are separately optimized.
The proposed methods closely achieve the global optimum obtained by the genetic algorithm, outperforming the state-of-the-art schemes.

The proposed approach focuses on designing a non-differentiable trajectory by integer programming. 
We underscore that applying the non-iterative approach to build a differentiable trajectory is another important open problem.
We attach the preliminary results on continuous network optimization conducted on the proposed environment in Appendix~\ref{appendix:Studies on the Continous Optimizations with Moving Users}.
Then continuous decision-making algorithms such as deep deterministic policy gradient \cite{LillicrapHPHETS15-ddpg} should be adopted.

Optimizing multi-UAV networks through the MDP formulation is another open challenge.
A proliferation of research on multi-UAV networks has adopted iterative optimization, and there might be a gap in network utility due to the curse of the initial choice.

We expect that the network utility of aerial networks can be improved by incorporating the non-iterative approach.


\begin{appendices}

\section{Related Works}
\label{Appendix:Related Works}
Recent works that suggest TP and resource management of UAV-BSs in various scenarios are categorized into the following subsections.
Key characteristics of the works are summarized in Table.~\ref{tab:related_works}.
\subsection{UAV-BSs in Time-Critical Mobile Scenarios}
Several research works have studied the control of UAV-BSs in time-critical wireless sensor networks \cite{Al-Hilo_RIS_2d-TP, Samir_TimeConstarined_IoT, Hu_AoI_MAC, Tong_DRL_TP_AoI}.
The works in \cite{Samir_TimeConstarined_IoT, Al-Hilo_RIS_2d-TP} target to maximize the number of devices that successfully transmit data within the lifetime of data.
In \cite{Samir_TimeConstarined_IoT}, two algorithms are suggested to jointly design a 2-dimensional trajectory and bandwidth allocation of devices.
Reconfigurable intelligent surfaces are additionally considered in \cite{Al-Hilo_RIS_2d-TP} and jointly optimized with the trajectory of a UAV-BS by using deep-reinforcement learning.
In \cite{Hu_AoI_MAC}, the authors propose an age of information minimization problem in a wireless sensor network, where sensors are powered by a UAV-BS using energy harvesting.
These works enhance the network utility for timely data, but the fairness between the users is not considered.
Without considering fairness, the trajectory of a UAV-BS can be biased according to the QoS constraints and channel state of users, thereby decreasing the number of served IoT users.

\subsection{UAV-BSs for Wireless Sensor Networks}
The work in \cite{You_TP_DataHarvesting} addresses a problem that maximizes the minimum average data rate in wireless sensor networks under an outage probability constraint and angle-dependent Rician fading.
In \cite{Zhan_TP_WSN}, the authors reformulate the TP problem into an equivalent traveling salesman problem to maximize the number of visited wireless sensors.
The authors of \cite{Zhu_TP_Recharge_station} suggest a TP algorithm that minimizes the required time slots for collecting data from all wireless sensors.
In \cite{Zhu_TP_DRL_Energy}, a deep-reinforcement learning model is adopted to minimize energy consumption while collecting data from wireless sensors.
These studies provide a well-designed trajectory in various mobile network scenarios, but designing optimal RA and PC still remains unsolved, which should be jointly designed with the trajectory.

\subsection{TP of UAV-BSs}
The works in \cite{Liu_TP_PC_sumrate_RL, Bayerlein_TP_sumrate_RL, Abbasi_TP_PC_sumrate_NOMA, Shen_multi_UAV_TP_PC_sumrate} propose a TP problem maximizing the sum-rate of users in UAV-enabled networks.
These works significantly enhance the sum-rate by adopting reinforcement-learning (RL) methods \cite{Liu_TP_PC_sumrate_RL, Bayerlein_TP_sumrate_RL} or utilizing successive convex approximation  \cite{Shen_multi_UAV_TP_PC_sumrate, Abbasi_TP_PC_sumrate_NOMA}.
However, none of these works considers the fairness between users or the RA problem.

The authors in \cite{Zeng_OFDMA_relay_GLOBECOM, Zeng_OFDMA_relay_TWC} propose a TP algorithm that jointly optimizes RA and PC parameters to maximize the fairness of users in the orthogonal frequency-division multiplexing access (OFDMA) network, where a single UAV-BS is utilized as a relay node.
These algorithms sequentially find the position, RA, and PC parameters by optimizing these parameters for the next time slot based on the current location of the UAV-BS.
The problems designed in these works are well-investigated with solid results, 
but future network states still need to be considered when optimizing the current movement of the UAV-BS.

\section{Proof of Proposition \ref{prop1:1} \label{appendix:prop1}}

The logarithm of the sum-rate of $i$-th user is reformulated as
    \begin{align} 
        \log R_i &= \log \sum_{t\in\mathcal{T}} R_i^{(t)} \\
        &= \log \prod_{t=1}^{T} \left(\frac{\sum_{k=0}^t R_i^{(k)}}{\sum_{k=0}^{t-1} R_i^{(k)}}\right)\cdot R_i^{(0)} \\
        &= \sum_{t=1}^{T} \log \left(1 + \frac{R_i^{(t)}}{\sum_{k=0}^{t-1} R_i^{(k)}}\right) + \log R_i^{(0)},
    \end{align}
    where the auxiliary constant $R_i^{(0)}$ is identical for all $i\in\mathcal{I}$.
    In what follows, we shall adopt the auxiliary constant $R_i^{(0)}$ to consistently decouple the problem for all time slots\footnote{Otherwise, we have a different term at $t=1$.}.
    For the numerical experiments, $R_i^{(0)},~i\in\mathcal{I}$ are chosen to be $1$, but the value of $R_i^{(0)}$ rarely affects on the outcome of the proposed algorithms.
    For each user $i$, $R_i^{(0)}$ determines the cumulative throughput $\sum_{k=0}^t R_i^{(k)}$, which affects the solution of RA and PC.
    A specific range of $R_0$ can affect the short-term RRM; but does not have a significant long-term effect, as the cumulative throughput naturally grows as time goes by.
    
    The objective in \eqref{eq:p1_objective_function} is equivalent with
    \begin{align}
        \sum_{i\in \mathbf{A}} \log R_i = \sum_{t=1}^{T} \sum_{i\in \mathbf{A}^{(t)}} \log \left(1 + \frac{R_i^{(t)}}{\sum_{k=0}^{t-1} R_i^{(k)}}\right) + \tilde{R},
    \end{align}
    where $\tilde{R} = \sum_{i\in \mathbf{A}} \log R_i^{(0)}$.
    Then, the following inequality holds:
    \begin{align}
        & \max_{\mathbf{q},\mathbf{A}, \mathbf{B},\mathbf{P}} \sum_{i\in \mathbf{A}} \log R_i \label{eq:p1_objective_reformulation}\\
        & = \max_{\mathbf{q},\mathbf{A}, \mathbf{B},\mathbf{P}} \sum_{t=1}^{T} \sum_{i\in \mathbf{A}^{(t)}} \log \left(1 + \frac{R_i^{(t)}}{\sum_{k=0}^{t-1} R_i^{(k)}}\right) + \tilde{R} \\
        & \geq \sum_{t=1}^{T}\max_{\mathbf{q}^{(t)},\mathbf{A}^{(t)}, \mathbf{B}^{(t)},\mathbf{P}^{(t)}} \sum_{i\in \mathbf{A}^{(t)}} \log \left(1 + \frac{R_i^{(t)}}{\sum_{k= 0}^{t-1} R_i^{(k)}}\right) .
        \label{eq:sum_pt}
    \end{align}
    The inequality \eqref{eq:sum_pt} indicates that the addition of the objectives of Problem $\bf{\mathdutchcal{P}^{(t)}}, t\in \mathcal{T}$ is the lower bound of Problem $\bf{\mathdutchcal{P}1}$.

\section{Derivation of the Initial \texorpdfstring{$\mathbf{B}^{(t)}$}{}}
\label{appendix:derive_beta}
If $\mathbf{A}^{(t)}$ is given, Problem \eqref{eq:p_UA_RA} is concave; thus can be solved by finding the KKT conditions.
For Problem \eqref{eq:p_UA_RA_objective}, the Lagrangian $\mathcal{L}_{\mathbf{B}|\mathbf{A}}(\cdot)$ is represented as
\begin{align}
    &\mathcal{L}_{\mathbf{B}|\mathbf{A}}(\mathbf{B}^{(t)}, \boldsymbol{\nu}^{(t)}, \lambda^{(t)})
    = \sum_{i\in \mathbf{A}^{(t)}} \log (1 + w_i^{(t)}\beta_i^{(t)}) \\
    &+ \sum_{i\in \mathbf{A}^{(t)}} \nu_i^{(t)}\bigg(\beta_i^{(t)}-\frac{\alpha_i^{(t)}r_i}{e_i^{(t)}}\bigg)
    - \lambda^{(t)}(\sum_{i\in \mathbf{A}^{(t)}}\beta_i^{(t)}-B),
\end{align}
where $w_i^{(t)}=\alpha_i^{(t)}d_i^{(t)}e_i^{(t)}/\sum_{k=0}^{t-1}R_i^{(k)}$; $\boldsymbol{\nu}^{(t)}=[\nu_i^{(t)}]_{i\in\mathcal{I}}$ and $\lambda^{(t)}$ are Lagrangian coefficient.

To meet the first-order optimality, we have
\begin{align}
    \nabla_{\beta_i^{(t)}} \mathcal{L}_{\mathbf{B}|\mathbf{A}}(\cdot) = \frac{w_i^{(t)}}{1+w_i^{(t)}\beta_i^{(t)}} + \nu_i^{(t)}-\lambda^{(t)} = 0.
    \label{UA_RA_first_optimality}
\end{align}
\begin{align}
    \nu_i^{(t)}\bigg(\beta_i^{(t)}-\frac{\alpha_i^{(t)}r_i}{e_i^{(t)}}\bigg) = 0.
    \label{UA_RA_slackness}
\end{align}
By substituting $\nu_i^{(t)}$ in \eqref{UA_RA_slackness} with \eqref{UA_RA_first_optimality}, the following inequality holds:
\begin{align}
    \bigg(\lambda^{(t)}-\frac{w_i^{(t)}}{1+w_i^{(t)}\beta_i^{(t)}} \bigg)\bigg(\beta_i^{(t)}-\frac{\alpha_i^{(t)}r_i}{e_i^{(t)}}\bigg)=0.
    \label{UA_RA_zero_condition}
\end{align}
For users $i\in\mathbf{A}^{(t)}$, combining \eqref{UA_RA_zero_condition} and the minimum RA requirements constraints \eqref{eq:p_UA_RA_beta} results in
\begin{equation}
    \beta_i^{(t)} = \max\bigg(\frac{r_i}{e_i^{(t)}}, \frac{1}{\lambda^{(t)}}-\frac{1}{w_i^{(t)}}\bigg).
\end{equation}

Because $\beta_i^{(t)}$ is simply determined to be zero for $i\notin\mathbf{A}^{(t)}$, we can conclude
\begin{align}
   \beta_i^{(t)} = 
   \begin{cases}
        \max\bigg(\frac{r_i}{e_i^{(t)}}, \frac{1}{\lambda^{(t)}}-\frac{1}{w_i^{(t)}}\bigg) & \text{if $\alpha_i^{(t)}d_i^{(t)}=1$,} \\
        \hfil 0 & \text{otherwise.}
   \end{cases}
\end{align}

\section{Optimal Resource Allocation}
\label{Appendix:Optimal Resource Allocation}

If $\mathbf{A}^{(t)}$ and $\mathbf{P}^{(t)}$ are fixed, the problem \eqref{eq:lookahead_value} is written as
\begin{subequations}
    \begin{alignat}{3}
        & \max_{\mathbf{B}^{(t)}} && 
        \sum_{i\in \mathbf{A}^{(t)}} \log \left(1 + \frac{R_i^{(t)}}{\sum_{k= 0}^{t-1} R_i^{(k)}}\right) 
        \label{eq:p_resource_objective}\\
        & \text{~~s.t.~}
          && \sum_{i \in \mathbf{A}^{(t)}} \beta_i^{(t)} \leq B, 
        \label{eq:p_resource_sum_resource}\\
        & && \sum_{i \in \mathbf{A}^{(t)}} \rho_i^{(t)}\beta_i^{(t)} \leq P,
        \label{eq:p_resource_sum_power}\\
        & && \beta_i^{(t)} \geq \alpha_i^{(t)}r_i/e_i^{(t)}, \forall i. 
        \label{eq:p_resource_combined_beta}
    \end{alignat}
    \label{eq:p_resource}
\end{subequations}
The optimal solution of the problem \eqref{eq:p_resource} can be obtained by using the Lagrangian dual method.
Then, $\mathbf{B}^{(t)}$ can be optimized by globally searching all feasible $\mathbf{B}^{(t)}$ for the objective function of \eqref{eq:p_resource}.

\begin{algorithm}[tb]
\caption{Resource Allocation}
\label{alg:user_resource}
    \KwInput{$\mathbf{q}^{(t)}$, $\mathbf{A}^{(t)}$, $\mathbf{P}^{(t)}$.}\\
    \KwOutput{$\mathbf{B}^{(t)}$.}\\
    \KwInitialize{$\boldsymbol{\mu}^{(t)}$, $\lambda_1^{(t)}$, $\lambda_2^{(t)}$, $\hat{\mathcal{L}}$ $\leftarrow 0$, $\hat{\mathcal{L}}_{\text{prev}}\leftarrow \infty$} \\
    \While{$\|\hat{\mathcal{L}}-\hat{\mathcal{L}}_{\mathrm{prev}}\| > \epsilon$}{
        Update $\boldsymbol{\mu}^{(t)}, \lambda_1^{(t)}, \lambda_2^{(t)}$ according to \eqref{eq:resource_lagrangian_update_mu}-\eqref{eq:resource_lagrangian_update_lambda_2}.\\
        $\hat{\mathcal{L}}_{\text{prev}} \leftarrow \hat{\mathcal{L}}$,~
        $\hat{\mathcal{L}} \leftarrow \mathcal{L}_\beta^*(\boldsymbol{\mu}^{(t)}, \lambda_1^{(t)}, \lambda_2^{(t)})$ using \eqref{eq:p_resource_dual_objective}. \\
    }
    $\mathbf{B}^{(t)} \leftarrow \left[\frac{1}{\lambda_1^{(t)}+\lambda_2^{(t)}\rho_i^{(t)}-\mu_i^{(t)}} - \frac{1}{w_i^{(t)}}\right]_{i\in\mathcal{I}}$. \\
\end{algorithm}
     
The Lagrangian dual problem of \eqref{eq:p_resource} is 
\begin{subequations}
    \begin{alignat}{3}
        & \min_{\boldsymbol{\mu}^{(t)}, \lambda_1^{(t)}, \lambda_2^{(t)}} && 
        \mathcal{L}_{\mathbf{B}}^*(\boldsymbol{\mu}^{(t)}, \lambda_1^{(t)}, \lambda_2^{(t)}) \\
        & \text{~~~~~~s.t.~}
          && \mu_i^{(t)} \geq 0, \forall i, \\
        & && \lambda_1^{(t)}, \lambda_2^{(t)} \geq 0,
    \end{alignat}
    \label{eq:p_resource_dual}
\end{subequations}
where $\boldsymbol{\mu}^{(t)}=[\mu_i^{(t)}]_{i\in\mathcal{I}}$ and 
\begin{align}
    \label{eq:p_resource_dual_objective}
    \mathcal{L}&_{\mathbf{B}}^*(\boldsymbol{\mu}^{(t)}, \lambda_1^{(t)}, \lambda_2^{(t)})  \nonumber \\
    =& - \sum_{i\in \mathbf{A}^{(t)}}\alpha_i^{(t)} 
    -\sum_{i\in \mathbf{A}^{(t)}} \log(\lambda_1^{(t)}+\lambda_2^{(t)}\rho_i^{(t)}-\mu_i^{(t)}) \nonumber  \\
    &- \sum_{i\in \mathbf{A}^{(t)}} \mu_i^{(t)}\bigg(\frac{1}{w_i^{(t)}} + \frac{\alpha_i^{(t)}r_i}{e_i^{(t)}}\bigg) \nonumber \\
    &+ \lambda_1^{(t)}\bigg(\sum_{i\in \mathbf{A}^{(t)}}\frac{1}{w_i^{(t)}}+B\bigg) 
    + \lambda_2^{(t)}\bigg(\sum_{i\in \mathbf{A}^{(t)}}\frac{\rho_i^{(t)}}{w_i^{(t)}}+P\bigg). 
\end{align}
The derivation of the Lagrangian dual function $\mathcal{L}_{\mathbf{B}}^*(\boldsymbol{\mu}^{(t)}, \lambda_1^{(t)}, \lambda_2^{(t)})$ is shown in the following section Appendix~\ref{appendix:derive_Lag}.
The duality gap between \eqref{eq:p_resource} and \eqref{eq:p_resource_dual} is zero, because the objective in \eqref{eq:p_resource_objective} is concave and the constraints \eqref{eq:p_resource_sum_resource}-\eqref{eq:p_resource_combined_beta} are linear.
Then, by using the sub-gradient descent method, the Lagrangian multipliers can be updated by 
\begin{align}
    & \mu_i^{(t)} \leftarrow \mu_i^{(t)} - \gamma\bigg(\frac{1}{o_i^{(t)}} 
    - \frac{1}{w_i^{(t)}} - \frac{\alpha_i^{(t)}r_i}{e_i^{(t)}}\bigg), 
    \label{eq:resource_lagrangian_update_mu}\\
    & \lambda_1^{(t)} \leftarrow \lambda_1^{(t)} - \gamma\bigg(- \sum_{i\in \mathbf{A}^{(t)}} \frac{1}{o_i^{(t)}}
    + B + \sum_{i\in \mathbf{A}^{(t)}} \frac{1}{w_i^{(t)}}\bigg), 
    \label{eq:resource_lagrangian_update_lambda_1}\\ 
    &\lambda_2^{(t)} \leftarrow \lambda_2^{(t)} - \gamma\bigg(- \sum_{i\in \mathbf{A}^{(t)}} \frac{\rho_i^{(t)}}{o_i^{(t)}}
    + P + \sum_{i\in \mathbf{A}^{(t)}} \frac{\rho_i^{(t)}}{w_i^{(t)}}\bigg),
    \label{eq:resource_lagrangian_update_lambda_2}
\end{align}
where $\gamma$ is a learning rate and $o_i^{(t)} = \lambda_1^{(t)}+\lambda_2^{(t)}\rho_i^{(t)}-\mu_i^{(t)}$.
The initial values of the multipliers are configured as $\mu_i^{(t)}=0.5$ for all $i\in \mathcal{I}$, $\lambda_1^{(t)}=0.5$, and $\lambda_2^{(t)}=0.8$.\footnote{We heuristically apply the bisection method to find the best combination that minimizes the convergence time.}
After that, the optimal value of $\mathbf{B}^{(t)}$ can be obtained from 
\begin{equation}
    \beta_i^{*(t)} = \frac{1}{\lambda_1^{(t)}+\lambda_2^{(t)}\rho_i^{(t)}-\mu_i^{(t)}} - \frac{1}{w_i^{(t)}},
\end{equation}
where $\mathbf{B}^{*(t)} = [\beta_i^{*(t)}]_{i\in\mathcal{I}}$.

\subsection{Derivation of the Lagrangian Dual Function \texorpdfstring{$\mathcal{L}_{\mathbf{B}}^*(\boldsymbol{\mu}^{(t)},~\lambda_1^{(t)}, \lambda_2^{(t)})$}{ } in \texorpdfstring{\eqref{eq:p_resource_dual_objective}}{(30)} 
\label{appendix:derive_Lag}}

The Lagrangian dual function \eqref{eq:p_resource_dual_objective} is derived as
\begin{align}
    \mathcal{L}_{\mathbf{B}}&(\mathbf{B}^{(t)}, \boldsymbol{\mu}^{(t)}, \lambda_1^{(t)}, \lambda_2^{(t)})  \\
    =& \sum_{i\in \mathbf{A}^{(t)}} \log (1 + w_i^{(t)}\beta_i^{(t)}) 
    + \sum_{i\in \mathbf{A}^{(t)}} \mu_i^{(t)}(\beta_i^{(t)}-\frac{\alpha_i^{(t)}r_i}{e_i^{(t)}}) \nonumber \\
    &- \lambda_1^{(t)}(\sum_{i\in \mathbf{A}^{(t)}}\beta_i^{(t)}-B) 
    - \lambda_2^{(t)}(\sum_{i\in \mathbf{A}^{(t)}}\rho_i^{(t)}\beta_i^{(t)}-P), \nonumber
\end{align}
where $w_i^{(t)} = \alpha_i^{(t)}d_i^{(t)}e_i^{(t)}/\sum_{k= 0}^{t-1} R_i^{(k)}$.

The partial derivative of $\mathcal{L}(\cdot)$ with respect to $\beta_i^{(t)}$ is 0 at optimal resource allocation vector $\mathbf{B}^{*(t)}$,
so we have
\begin{align}
    \nabla_{\beta_i^{(t)}} \mathcal{L}_{\mathbf{B}}(\cdot) &= \frac{w_i^{(t)}}{1+w_i^{(t)}\beta_i^{(t)}} + \mu_i^{(t)}-\lambda_1^{(t)}-\lambda_2^{(t)}\rho_i^{(t)} \\
    &= 0,
\end{align}
Then, the optimal $\beta_i^{*(t)}$ is
\begin{equation}
    \beta_i^{*(t)} = \frac{1}{\lambda_1^{(t)}+\lambda_2^{(t)}\rho_i^{(t)}-\mu_i^{(t)}} - \frac{1}{w_i^{(t)}},
\end{equation}
and $\mathbf{B}^{*(t)} = [\beta_i^{*(t)}]_{i\in\mathcal{I}}$.
Therefore, we have
\begin{align}
     \mathcal{L}&_{\mathbf{B}}^*(\boldsymbol{\mu}^{(t)}, \lambda_1^{(t)}, \lambda_2^{(t)})  
    = \mathcal{L}_\beta(\mathbf{B}^{*(t)}, \boldsymbol{\mu}^{(t)}, \lambda_1^{(t)}, \lambda_2^{(t)}).
\end{align}


\section{Optimal Power Control}
\label{Appendix:Optimal Power Control}
If $\mathbf{A}^{(t)}$ and $\mathbf{B}^{(t)}$ are given, the problem \eqref{eq:lookahead_value} is written as
\begin{subequations}
    \begin{alignat}{3}
        & \max_{\mathbf{P}^{(t)}} && 
        \sum_{i\in \mathbf{A}^{(t)}} \log \left(1 + \frac{R_i^{(t)}}{\sum_{k= 0}^{t-1} R_i^{(k)}}\right) 
        \label{eq:p_power_objective}\\
        & \text{~~s.t.~}
          && \rho_i^{(t)} \geq 0, \forall i, 
        \label{eq:p_power_psd}\\
        & && \sum_{i \in \mathbf{A}^{(t)}} \rho_i^{(t)}\beta_i^{(t)} \leq P, 
        \label{eq:p_power_sum}\\
        & && R_i^{(t)}\geq \alpha_i^{(t)}r_i, \forall i.
        \label{eq:p_power_user_QoS}
    \end{alignat}
    \label{eq:p_power}
\end{subequations}
Since $\log(1+x) \approx x$ when $x\ll1$, we relax the objective function as
\begin{equation}
    \sum_{i\in \mathbf{A}^{(t)}} \log \left(1 + \frac{R_i^{(t)}}{\sum_{k= 0}^{t-1} R_i^{(k)}}\right)
    \approx \sum_{i\in \mathbf{A}^{(t)}}  \frac{R_i^{(t)}}{\sum_{k= 0}^{t-1} R_i^{(k)}}.
    \label{eq:p_power_approx_objective} 
\end{equation}
We remark that the above approximation holds more accurately as $t$ increases because the cumulative received data $\sum_{k= 0}^{t-1} R_i^{(k)}$ increases over time.
Also, the constraint \eqref{eq:p_power_user_QoS} can be equivalently expressed with respect to the PSD of a user as follows:
\begin{equation}
    \rho_i^{(t)} 
    \geq \zeta_i^{(t)} , \forall i,
    \label{eq:p_power_user_QoS_wrt_psd}
\end{equation}
where $\zeta_i^{(t)} = \frac{N_0}{10^{-\xi_i^{(t)}/10}}\big(2^{\alpha_i^{(t)}r_i/\beta_i^{(t)}}-1\big)$.
We can replace \eqref{eq:p_power_psd} and \eqref{eq:p_power_user_QoS} by \eqref{eq:p_power_user_QoS_wrt_psd} because $\zeta_i^{(t)} \geq 0$.

By applying \eqref{eq:p_power_approx_objective} and \eqref{eq:p_power_user_QoS_wrt_psd}, the problem \eqref{eq:p_power} can be reformulated as
\begin{subequations}
    \begin{alignat}{3}
        & \max_{\mathbf{P}^{(t)}} && 
        \sum_{i\in \mathbf{A}^{(t)}} \frac{R_i^{(t)}}{\sum_{k= 0}^{t-1} R_i^{(k)}}
        \label{eq:p_approx_power_objective}\\
        & \text{~~s.t.~}
          && \sum_{i \in \mathbf{A}^{(t)}} \rho_i^{(t)}\beta_i^{(t)} \leq P, 
        \label{eq:p_approx_power_sum}\\
        & && \rho_i^{(t)} \geq \zeta_i^{(t)}, \forall i.
        \label{eq:p_approx_power_user_QoS_wrt_psd}
    \end{alignat}
    \label{eq:p_approx_power}
\end{subequations}
The objective function and constraints of \eqref{eq:p_approx_power} are concave and linear, respectively.
Therefore, the zero-duality gap is guaranteed between Problem \eqref{eq:p_approx_power} and its dual problem.

The Lagrangian $\mathcal{L}_{\mathbf{P}}$ of \eqref{eq:p_approx_power} is
\begin{align}
    \label{eq:lagrangian_power}
    \mathcal{L}&_{\mathbf{P}}(\mathbf{P}^{(t)}, \boldsymbol{\nu}^{(t)}, \lambda^{(t)})
    = \sum_{i\in \mathbf{A}^{(t)}} \tau_i^{(t)}\log_2(1 + \omega_i^{(t)}\rho_i^{(t)}) \nonumber \\
    &+ \sum_{i\in \mathbf{A}^{(t)}} \nu_i^{(t)}(\rho_i^{(t)}-\zeta_i^{(t)})
    - \lambda^{(t)}\big(\sum_{i\in \mathbf{A}^{(t)}}\rho_i^{(t)}\beta_i^{(t)}-P\big), 
\end{align}
where $\boldsymbol{\nu}^{(t)}=[\nu_i^{(t)}]_{i\in \mathcal{I}}$, $\tau_i^{(t)} = \alpha_i^{(t)}d_i^{(t)} /\sum_{k= 0}^{t-1} R_i^{(k)}$ and 
$\omega_i^{(t)} = 10^{\xi_i^{(t)}/10}/n_0$.

We need to find a point at which derivative of $\mathcal{L}_{\mathbf{P}}(\cdot)$ is zero:
\begin{align}
    \nabla_{\rho_i^{(t)}} \mathcal{L}_{\mathbf{P}}(\cdot) = \frac{\tau_i^{(t)}\omega_i^{(t)}}{1+\omega_i^{(t)}\rho_i^{(t)}} + \nu_i^{(t)}-\lambda^{(t)}\beta_i^{(t)}
    = 0.
\end{align}
In addition, the pair of $\mathbf{P}^{(t)}, \boldsymbol{\nu}^{(t)}$ and $\lambda^{(t)}$ should satisfy
\begin{equation}
    \nu_i^{(t)}(\rho_i^{(t)}-\zeta_i^{(t)}) = 0
\end{equation}
to meet the complementary slackness of a KKT solution.

Similar to Appendix~\ref{appendix:derive_beta}, $\rho_i^{(t)}$ is determined as
\begin{align}
    \hspace{-.1cm}
    \rho_i^{(t)} = 
    \begin{cases}
        \max\bigg(\zeta_i^{(t)}, \frac{\tau_i^{(t)}}{\beta_i^{(t)}\lambda^{(t)}}-\frac{1}{\omega_i^{(t)}}\bigg) & \text{if $\alpha_i^{(t)}d_i^{(t)}=1$,} \\
        \hfil 0 & \text{otherwise.}
    \end{cases}
    \label{p:rho_solution}
\end{align}

\section{
DQN Architecture and Training Procedure
}
\label{appendix:dqn}

\begin{table}[htb]
    \centering
    \caption{DQN Parameter Configurations}
    \begin{tabular}{cc}
    \toprule
    Parameter & Value \\ \cmidrule(r){1-1}\cmidrule{2-2}
    Learning rate (LR) & $1 \cdot 10^{-4}$ \\
    LR scheduler step size & 500 \\
    LR decay factor & 0.9 \\
    Discount factor & 0.99 \\
    Exploration prob. $\epsilon$ & 1 \\
    $\epsilon$-decay per update & 0.999 \\
    Soft-update weight $\tau$ & $10^{-3}$ \\
    Soft-update period & 1 \\
    Experience-replay buffer size & $10^5$ \\
    Experience-replay batch size & $512$ \\
    \bottomrule
    \end{tabular}
    \label{table:DQN_parameters}
    \vspace{-10pt}
\end{table}

We design a 16-layer fully connected network to produce the results in Sec.~\ref{sec:Simulation_Results}.
All layers, except for the input and output layers, have input and output vectors of size $\mathbb{R}^{1024}$.
The input vector size is a cardinality of a state vector, determined as $11+8I$ by the number of users $I$.
The output vector size is $\mathbb{R}^{7}$ for our numerical experiment, determined by the cardinality of $S(\cdot)$.
A detailed configuration is provided on the implementation source code.

\begin{figure}[htb]
    \centering
    \hfill
    \subfloat{\includegraphics[width=\columnwidth]{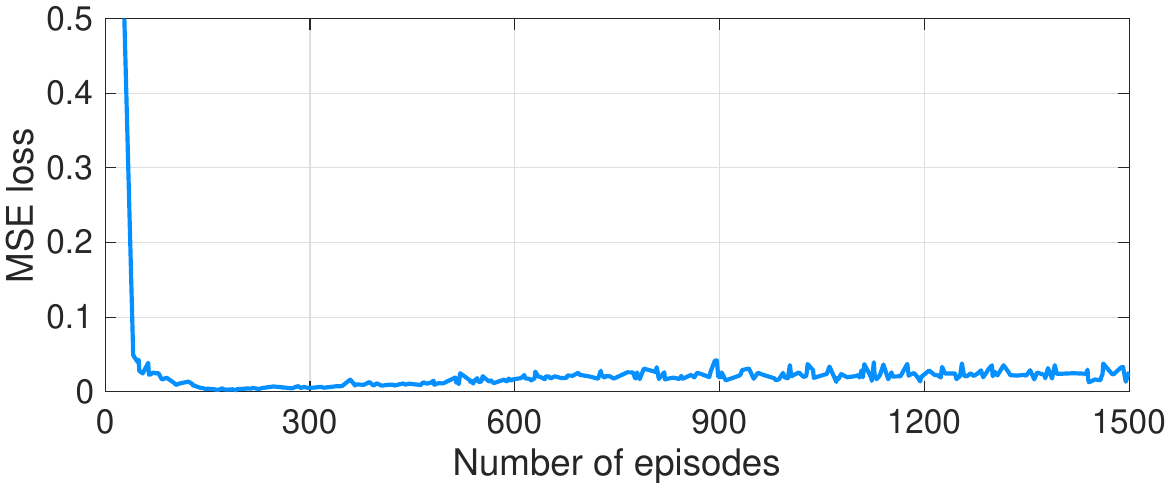}}
    \hfill
    \medskip
    \hfill
    \subfloat{\includegraphics[width=\columnwidth]{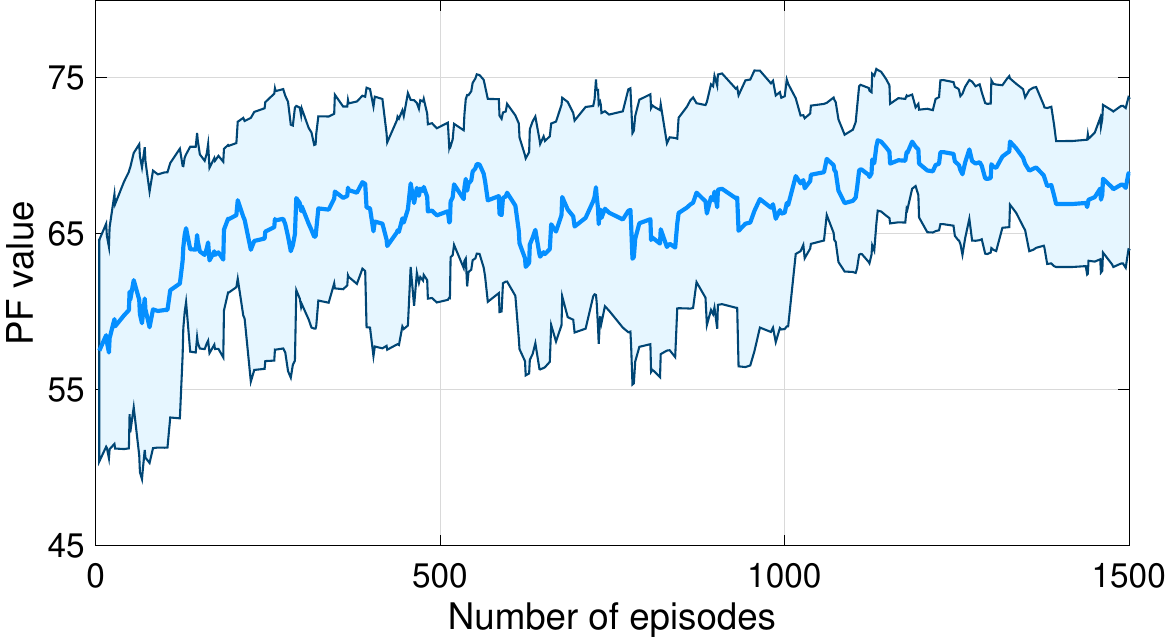}}
    \caption{
    Loss and PF value change for the first 1500 training episodes. 
    The PF value is moving-averaged with the window size 13. The colored area represents the variance.
    }
    \label{fig:dqn_training}
\end{figure}

The model is trained using the adaptive moment estimation (Adam) optimizer \cite{Kingma15-ADAM_optimizer}, step-decaying learning rate scheduler (e.g. StepLR in PyTorch), mean squared error (MSE) loss. 
All hyper-parameters are listed in Table~\ref{table:DQN_parameters}.
The DQN models are trained during 5,000 episodes, which is equivalent to 200,000 steps.
Figure~\ref{fig:dqn_training} illustrates the training progress 
with the DQN environment configured as 10 MHz bandwidth; 20 users; and $r_i=5$ for all $i\in \mathcal{I}$.

In all cases, the DQN scheme shows comparable results with the GA-TP and DFS but fails to exceed the PF of the DFS scheme.
The main reason is that the DFS scheme benefits from the analytical optimal reward of the sub-trajectory, but the DQN scheme should estimate the exact reward of each position $\mathbf{q}^{(t)}$ for all $t\in\mathcal{T}$.

\begin{figure*}[htb]
    \centering
    \null\hfill
    \subfloat[Discrete trajectory obtained from GA-TP.]{\includegraphics[width = .40\textwidth]{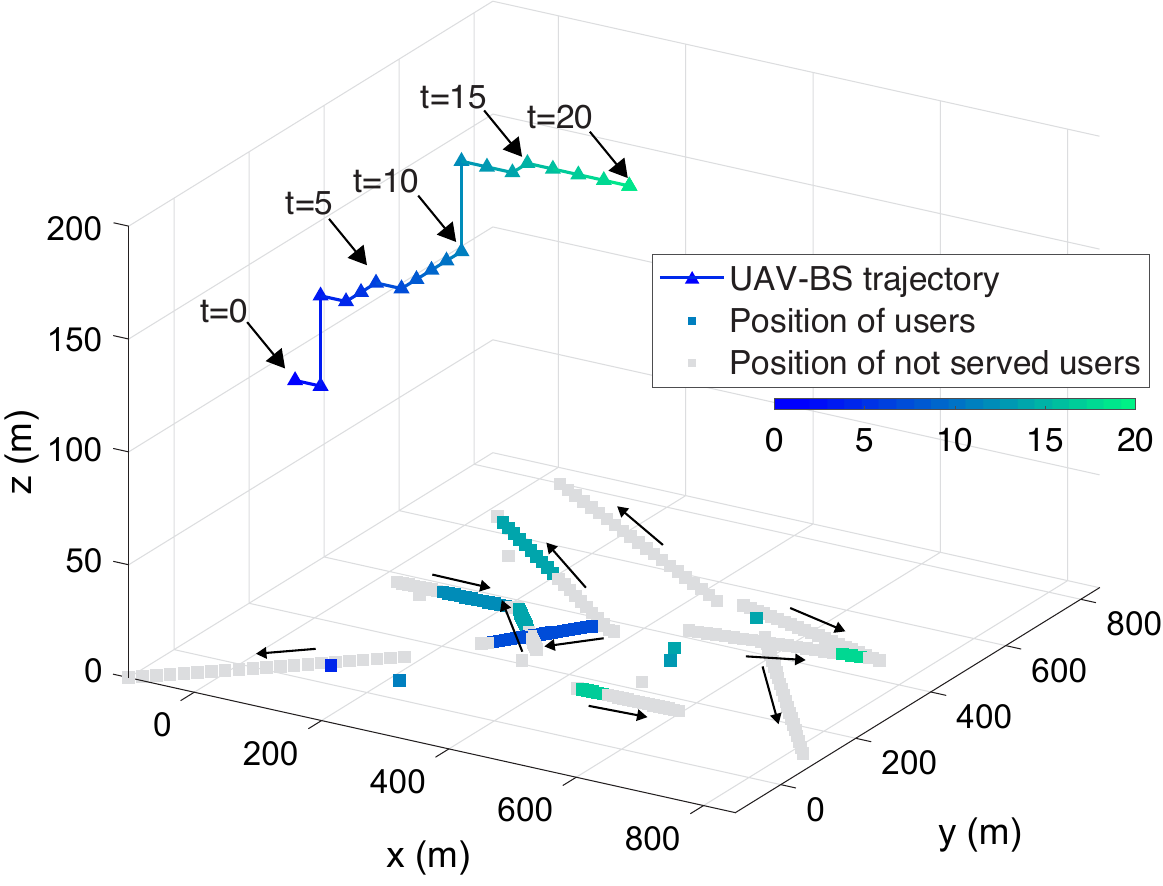}}
    \hfill
    \subfloat[Trajectory interpolated by the cubic spline.]{\includegraphics[width = .40\textwidth]{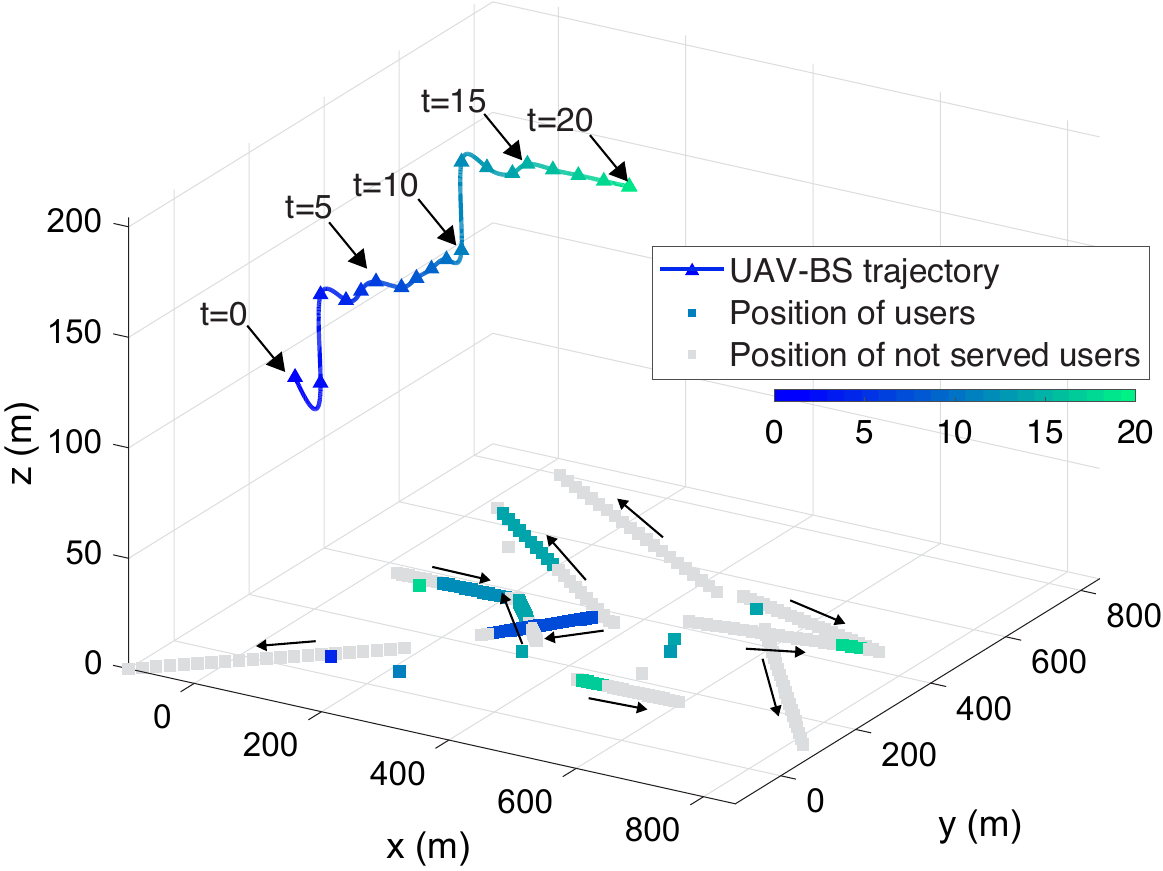}}
    \hfill\null
    \caption{
    Trajectories optimized with 10 moving users and 10 stationary users.
    }
    \label{Fig:trajectory visualization with mobility}
\end{figure*}
\begin{figure*}[htb]
    \centering
    \null\hfill
    \subfloat{\includegraphics[width = .35\textwidth, valign=c]{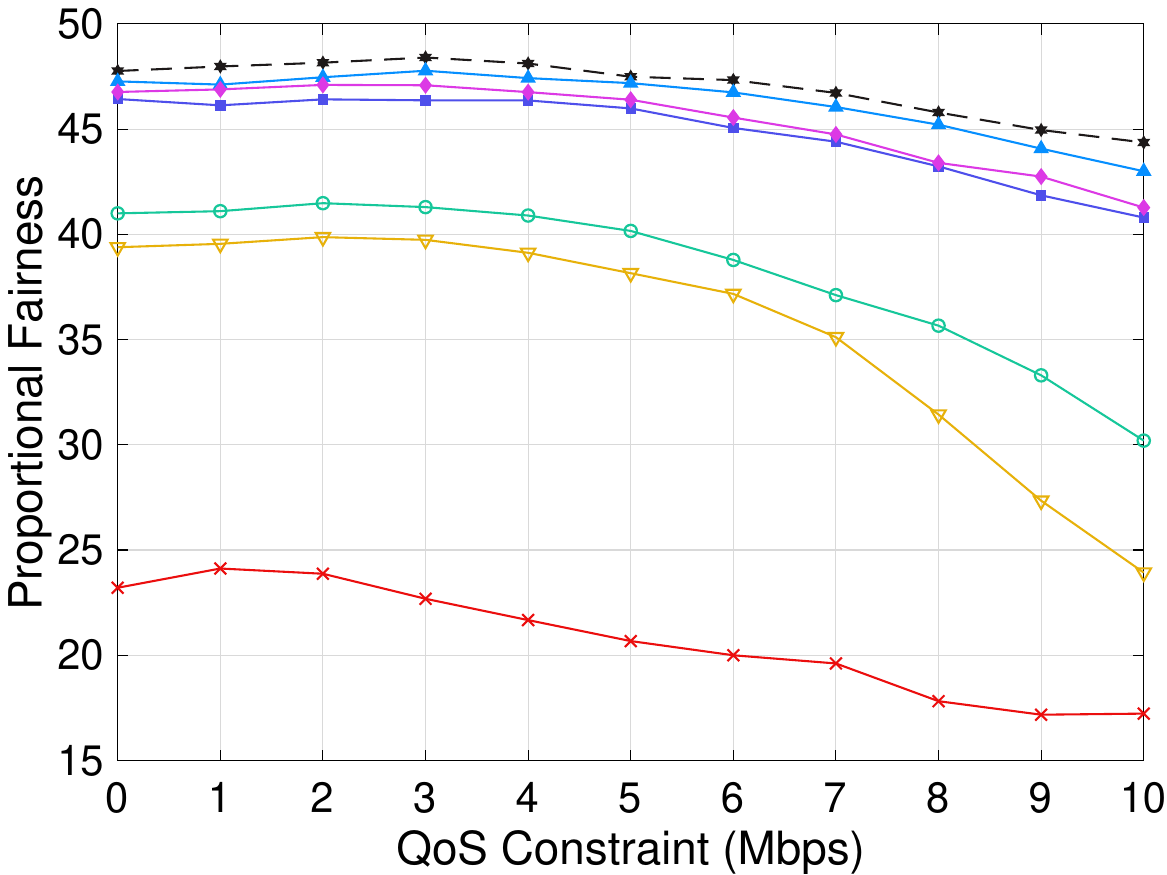}}
    \hfill
    \subfloat{\includegraphics[width = .35\textwidth, valign=c]{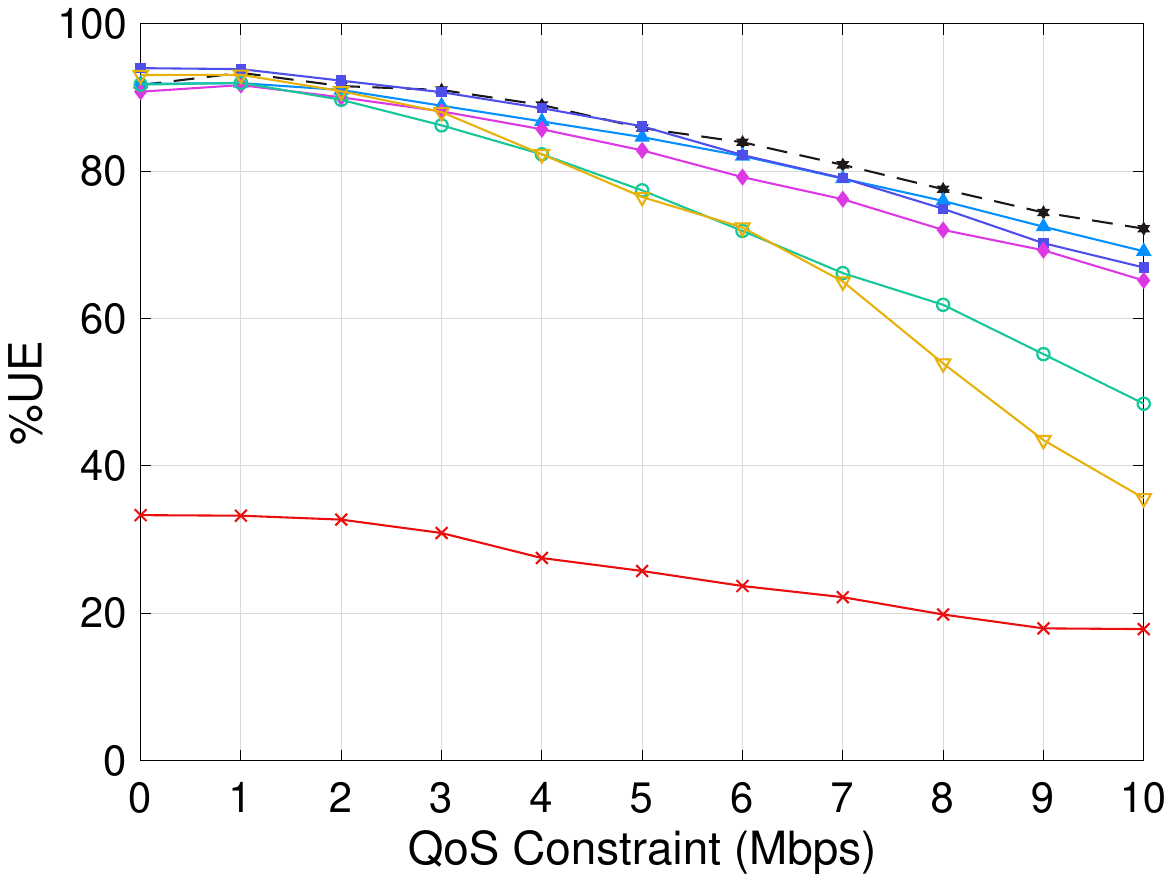}}
    \hfill
    \subfloat{\includegraphics[width = .14\textwidth, valign=c]{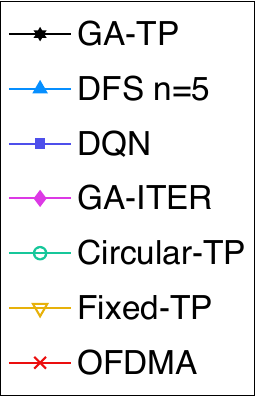}}
    \hfill\null
    \caption{
    Proportional fairness and number of served users with various QoS constraints in the dynamic environment.
    }
    \label{Fig:pf and coverage with mobility}
\end{figure*}

\section{
Studies on the Continous Optimizations with Moving Users
}
\label{appendix:Studies on the Continous Optimizations with Moving Users}

We additionally design an continuous TP and RRM to verify how the proposed method operates in the realistic scenario.
In this scenario, we additionally apply a uniform velcocity to half of the mobile users, considering future developments where IoT devices, such as autonomous cars, AAM/UAMs, and quadrupedal robots, will incorporate mobility. In the following experiments, we assume 10 users are mobile while the other 10 remain
stationary.\footnote{The velocity of users is randomly assigned with the speed ranging from $[0, 30]$ m/s.}

We first design a differentiable trajectory by interpolating the result trajectory obtained from the DFS method as shown in Fig.~\ref{Fig:trajectory visualization with mobility}.\footnote{The trajectory is interpolated using cubic splines.}
Then, we measured the PF in two cases: i) reusing the RRM obtained from the DFS algorithm (continuous trajectory w/o RRM); and ii) performing the RRM every 10 ms (continuous trajectory with RRM).

\begin{figure}[htb]
    \centering
    \includegraphics[width=0.47\textwidth]{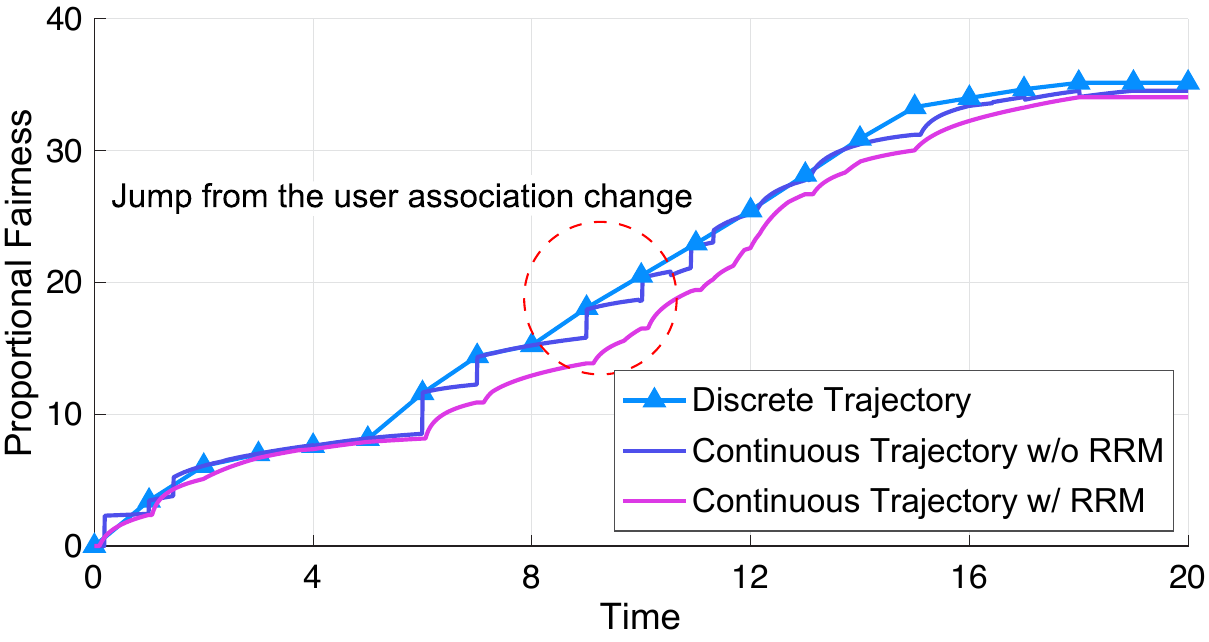}
    \caption{The proportional fairness per time slot for a single episode.}
    \label{fig:graph_env_0002-depth_5-ue_20}
\end{figure}

Fig.~\ref{Fig:pf and coverage with mobility} indicate the PF and \%UE over QoS constratins in the above environmental configurations.
There is no significant change in trend compared to Figs.~\ref{fig:pf-rate} and~\ref{fig:coverage-rate}.
However, slight decreases---1-2 for PF and 2-4\% for \%UE---are observed.

By making the TP and RRM continuous, the resulting PF measurement also becomes continuous as in Fig.~\ref{fig:graph_env_0002-depth_5-ue_20}.
For ``Continuous Trajectory without RRM", we just apply the RRM of the Discrete Trajectory---$\mathbf{A}^{(t)}$, $\mathbf{B}^{(t)}$, and $\mathbf{P}^{(t)}$ for $t\in\mathcal{T}$---to compute the PF at time $t'\in [t, t+1)$.
Then, we can observe multiple discrete jumps in PF as the user association discretely changes at every $t$-th time slot.
Thus, these discrete jumps disappear as the user association changes more granularly when we re-optimize the RRM every 10 ms for the continuous trajectory in ``Continuous Trajectory with RRM".

\section{Wall-Clock Time Benchmarks of the TP Algorithms}
\label{appendix:Wall-Clock Time Benchmarks of the TP Algorithms}

\begin{table}[htb]
    \centering
    \caption{Wall-clock time of TP algorithms}
    \label{Tab:Wall-clock time of TP algorithms}
    \begin{adjustbox}{width=\linewidth}
    \begin{tabular}{cccccccc}
    \toprule
    \multirow{2}{*}{Methods} & \multirow{2}{*}{GA-TP} & \multicolumn{3}{c}{DFS} & \multirow{2}{*}{DQN} & \multirow{2}{*}{GA-ITER} & \multirow{2}{*}{OFDMA} \vspace{-.03cm} \\
    \cmidrule(r){3-5}
    \vspace{-.05cm}
                            &                        & $n=1$  & $n=3$  & $n=5$ &                      &                          &                        \\
    \cmidrule(r){1-1} \cmidrule{2-8}
    Time (s)                & 84.73                  & 0.09   & 1.34   & 33.34 & 0.50                 & 22.33                    &      -                  \\
    PF                      & 48.86                  & 46.73  & 47.97  & 48.57 & 47.48                & 47.85                    & 22.10                 
    \\
    \bottomrule
    \end{tabular}
    \end{adjustbox}
\end{table}

Table \ref{Tab:Wall-clock time of TP algorithms} shows the wall-clock time benchmarks when a UAV-BS serves users with 2 MHz bandwidth during 20 time slots (which is equivalent to 60 seconds).
We omit the elapsed time of OFDMA because we implement the scheme using the genetic algorithm \cite{weise2009-genetic}, not as described in the original article \cite{Zeng_OFDMA_relay_TWC}.

The proposed schemes, DFS and DQN, demonstrate significantly faster path design capabilities while achieving higher PF values, compared to GA-ITER (which is an iterative framework).
This can be attributed to the fact that the proposed methods avoid ``the curse of initialization" without the need for repetitive optimization processes.

\end{appendices}


\bibliographystyle{IEEEtran}
\bibliography{references.bib}

\end{document}